\documentclass[11pt,a4paper]{article}
\usepackage[utf8]{inputenc}
\usepackage[english]{babel}
\usepackage{amsmath}
\usepackage{amsthm}
\numberwithin{equation}{section}
\usepackage{amsfonts}
\usepackage{mathtools}
\usepackage{mathrsfs}
\usepackage{mathpazo}
\usepackage{bbm}
\usepackage{amssymb}
\usepackage{xcolor}
\usepackage{graphicx}
\usepackage{braket}
\usepackage{bm}
\newcommand{\changefont}[3]{
\fontfamily{#1} \fontseries{#2} \fontshape{#3} \selectfont}
\changefont{ppl}{m}{n}

\newcommand{\llrrbracket}[1]{% \llrrbracket{..}
  \left[\mkern-3mu\left[#1\right]\mkern-3mu\right]}
\newcommand{\llrrparen}[1]{% \llrrparen{..}
  \left(\mkern-4mu\left(#1\right)\mkern-4mu\right)}

\newcommand{\bq}{\mathbf{q}}

\newcommand{\bE}{\mathbf{E}}
\newcommand{\bF}{\mathbf{F}}
\newcommand{\bZ}{\mathbf{Z}}
\newcommand{\bz}{\mathbf{z}}
\newcommand{\bfeta}{\bm{\eta}}

\newcommand{\bPhi}{\bm{\Phi}}

\newcommand{\PB}[2]{\left\lbrace #1 , #2 \right\rbrace}

\newcommand{\comment}[1]{\textit{ \color{gray}}}

\usepackage{a4wide}

\usepackage[bookmarks=true,colorlinks=true,linkcolor=black,citecolor=black,urlcolor=black,bookmarksnumbered]{hyperref}
\usepackage{cite}

\begin{document}

\setcounter{equation}{0}
\setcounter{footnote}{0}
\setcounter{section}{0}

\thispagestyle{empty}

\begin{flushright}\end{flushright}

\begin{center}
\vspace{1.5truecm}

%{\Large \bf World-volume currents in $E_{d(d)}$ generalised geometry for $d\leq 6$}

%{\Large \bf Currents, charges and algebras in $E_{d(d)}$ generalised geometry for $d\leq 6$}

{\Large \bf Currents, charges and algebras in exceptional generalised geometry}
\vspace{1.5truecm}

{David Osten}

\vspace{1.0truecm}

{\em Institute for Theoretical and Mathematical Physics \\
Lomonosov Moscow State University \\
Lomonosovsky Avenue, Moscow, 119991, Russia}

\vspace{1.0truecm}

{{\tt ostend@itmp.msu.ru}}

\vspace{1.0truecm}
\end{center}

\begin{abstract}
A classical $E_{d(d)}$-invariant Hamiltonian formulation of world-volume theories of half-BPS $p$-branes in type IIb and eleven-dimensional supergravity is proposed, extending known results to $d \leq 6$. It consists of a Hamiltonian, characterised by a generalised metric, and a current algebra constructed s.t. it reproduces the $E_{d(d)}$ generalised Lie derivative. $E_{d(d)}$-covariance necessitates the introduction of so-called charges, specifying the type of $p$-brane and the choice of section. For $p>2$, currents of $p$-branes are generically non-geometric due to the imposition of $U$-duality, e.g. the M5-currents contain coordinates associated to the M2-momentum.

A derivation of the $E_{d(d)}$-invariant current algebra from a canonical Poisson structure is in general not possible. At most, one can derive a current algebra associated to para-Hermitian exceptional geometry.

The membrane in the SL$(5)$-theory is studied in detail. It is shown that in a generalised frame the current algebra is twisted by the generalised fluxes. As a consistency check, the double dimensional reduction from membranes in M-theory to strings in type IIa string theory is performed. Many features generalise to $p$-branes in SL$(p+3)$ generalised geometries that form building blocks for the $E_{d(d)}$-invariant currents.
\end{abstract}

\pagebreak
\thispagestyle{empty}
\tableofcontents

\section{Introduction} 
O$(d,d)$ generalised geometry and double field theory \cite{Duff:1989tf,Tseytlin:1990nb,Tseytlin:1990va,Siegel:1993bj,Siegel:1993th,Hitchin:2004ut,Gualtieri:2003dx,Grana:2008yw,Hull:2009mi,Zwiebach:2011rg,Aldazabal:2013sca,Hohm:2013bwa,Geissbuhler:2013uka,Plauschinn:2018wbo} has been very useful in the study of world-sheet string theory in features like non-abelian and Poisson-Lie $T$-duality or classical world-sheet integrability \cite{Klimcik:1995dy,Klimcik:1996nq,Osten:2016dvf,Hassler:2017yza,Sakamoto:2017cpu,Borsato:2017qsx,Fernandez-Melgarejo:2017oyu,Sakamoto:2018krs,Lust:2018jsx,Demulder:2018lmj,Osten:2019ayq,Thompson:2019ipl,Borsato:2021gma}. In both cases arranging the degrees of freedom O$(d,d)$-covariantly was advantageous. For instance, a Hamiltonian formulation of strings in generic backgrounds is possible, using only input from generalised geometry: the generalised metric determines the Hamiltonian and the invariant O$(d,d)$-metric the Poisson structure \cite{Blair:2013noa,Osten:2019ayq}. In case of generalised parallelisable spaces, this formulation has been called $\mathcal{E}$-model and appeared first in the study of Poisson-Lie $T$-duality \cite{Klimcik:1995dy,Klimcik:1996nq,Sfetsos:1997pi,Klimcik:2015gba}.

At the same time, exceptional field theory and $E_{d(d)}$ generalised geometry \cite{Pacheco:2008ps,Berman:2010is,Berman:2011jh,Coimbra:2011ky,Berman:2012vc,Berman:2012uy,Coimbra:2012af,Hohm:2013pua,Lee:2014mla,Hohm:2014qga,Berman:2020tqn} has been a powerful tool in organising compactifications of eleven-dimensional supergravity, but not necessarily in understanding the underlying world-volume theories better. Additional complications in phrasing world-volume theories in an $E_{d(d)}$-invariant way are to be expected. That is, because $U$-duality acts, in general, non-perturbatively and maps different world-volume theories to each other -- in contrast to O$(d,d)$ that maps a string theory to a (not necessarily equivalent) string theory. For example, problems concerning $U$-duality of the classical membrane have been studied in \cite{Duff:1990hn,Duff:2015jka}.

The problem of $E_{d(d)}$-invariant world-volume formulations has, so far, mainly been studied at the level of construction of actions \cite{Sakatani:2016sko,Blair:2017hhy,Sakatani:2017vbd,Arvanitakis:2018hfn,Blair:2019tww,Sakatani:2020umt}. In all of these proposals the introduction of \textit{charges}, describing the kind of object one works with, was necessary in order to achieve duality covariance.

Success in the strive for an $\mathcal{E}$-model like formulation has, by now, only been achieved for the $E_{3(3)}$-theory and partly the $E_{4(4)}$-theory \cite{Duff:1990hn,Berman:2010is,Hatsuda:2012vm,Duff:2015jka,Sakatani:2020iad,Sakatani:2020wah}. Generalising \cite{Sakatani:2020iad}, such a formulation might -- similar to Poisson-Lie $T$-duality as canonical transformation of classical string theory -- help to understand to which extent the recently discussed Poisson-Lie $U$-duality, transformations of exceptional Drinfeld algebras or other interesting flux configurations \cite{duBosque:2017dfc,Ashmore:2018npi,Bakhmatov:2019dow,Sakatani:2019zrs,Malek:2019xrf,Sakatani:2020iad,Bakhmatov:2020kul,Blair:2020ndg,Musaev:2020bwm,Malek:2020hpo,Sakatani:2020wah,Gubarev:2020ydf,Musaev:2020nrt}, can be understood as some kind of classical duality transformation of world-volume theories in exceptional field theory.

The aim of this paper is to close that gap and propose an $E_{d(d)}$-covariant Hamiltonian formulation of world-volume theories, building on previous attempts \cite{Hatsuda:2012uk,Hatsuda:2012vm,Hatsuda:2013dya} and most recently \cite{Hatsuda:2020buq}. Besides the typical objects of $E_{d(d)}$ generalised geometry an additional object, the above mentioned charge, that specifies the type of world-volume, needs to be introduced to ensure $E_{d(d)}$-covariance.

Before an overview over the results is given, the string and its Hamiltonian formulation in terms of O$(d,d)$ generalised geometry is presented, in order to clarify the key points one might strive for in a generalisation to the $E_{d(d)}$-case.

\subsection{The string case}
The classical dynamics of a string coupling to the NS-NS part of an arbitrary supergravity background can be described by the Hamiltonian
\begin{equation}
H = \frac{1}{2} \int \mathrm{d} \sigma \mathcal{H}^{IJ} \big(G(x(\sigma)),B(x(\sigma))\big) \bE_I(\sigma) \bE_J(\sigma),
\end{equation}
where the embedding of the string into target space $x(\sigma)$ and the canonical momentum $p(\sigma)$ are collected in $\bE_M(\sigma) = (p_\mu(\sigma) , \partial x^\mu(\sigma))$, from now called \textit{current}. $\sigma$ is the spatial world-volume coordinate, $\partial = \partial_\sigma$. $\mathcal{H}$ is the generalised metric, encoding the $\sigma$-model couplings, metric $G$ and the $B$-field, in the Hamiltonian and the indices $I,J,K,...$ are raised and lowered by the O$(d,d)$-metric $\eta$:
\begin{equation}
\mathcal{H} = \left( \begin{array}{cc} G - BG^{-1}B & BG^{-1} \\ -G^{-1} B & G^{-1} \end{array} \right), \qquad \eta = \left( \begin{array}{cc} 0  & \mathbb{1} \\ \mathbb{1} & 0 \end{array} \right).
\end{equation}
The Virasoro constraints take the form $\mathcal{H}^{IJ} \bE_I \bE_J = 0 = \eta^{IJ} \bE_I \bE_J.$
\paragraph{Current algebra and generalised Lie brackets.} For a Hamiltonian description we still need to specify the Poisson brackets of the currents $\bE_I$ -- the \textit{current algebra}. In this article, we often discuss two types of current algebra brackets, the \textit{Dorfman current algebra}
\begin{equation}
\PB{\bE_I(\sigma)}{\bE_J(\sigma^\prime)}_D = \eta_{IJ} \partial \delta(\sigma - \sigma^\prime)\label{eq:STRINGCurrentAlgebraDorfman}
\end{equation}
and the \textit{Lie current algebra}
 \begin{equation}
\PB{\bE_I(\sigma)}{\bE_J(\sigma^\prime)}_L = \eta_{IJ} \frac{1}{2} (\partial -\partial^\prime)\delta(\sigma - \sigma^\prime) +  \omega_{IJ} \frac{1}{2} (\partial + \partial^\prime) \delta(\sigma - \sigma^\prime) \label{eq:STRINGCurrentAlgebraLie}
\end{equation}
where $\omega_{IJ} =  \left( \begin{array}{cc} 0  & - \mathbb{1} \\ \mathbb{1} & 0 \end{array} \right)$. The difference between these current algebra brackets is a total world-sheet derivative term  $\int \mathrm{d} \sigma \ \partial(...)$. Depending on the boundary conditions on $x(\sigma)$ such total derivative terms could be neglected, they correspond to boundary contributions for the open string and winding contributions for the closed string. They have been shown to be important on the quantum level \cite{Berman:2007vi} and, on the level of action, the difference between the O$(d,d)$-invariant version and the standard non-linear $\sigma$-model is a topological term \cite{Tseytlin:1990nb,Hull:2004in,Hull:2006qs}.

As functions on the phase space we consider functionals
\begin{equation}
\phi[x] = - \int\mathrm{d}\sigma \phi^I\big(x(\sigma)\big) \bE_I(\sigma) \quad \in (T\oplus T^\star)LM .
\end{equation}
The Poisson brackets of these reproduce the generalised Lie derivative of O$(d,d)$ generalised geometry for \eqref{eq:STRINGCurrentAlgebraDorfman}
respectively the canonical Lie bracket for \eqref{eq:STRINGCurrentAlgebraLie}:
\begin{equation}
\PB{\phi_1}{\phi_2}_D = [v_1 , v_2] + \mathcal{L}_{v_1} \xi_2 - \mathrm{d} (\iota_{v_2} \xi_1), \qquad \PB{\phi_1}{\phi_2}_L = [v_1 , v_2] + \mathcal{L}_{v_1} \xi_2 - \mathcal{L}_{v_2} \xi_1
\end{equation}
with $\phi_i = v_i + \xi_i \in (T\oplus T^\star)LM$. A detailed discussion can be found in \cite{Osten:2019ayq} and will be revisited in section \ref{chap:ParaHermitian}.

\paragraph{Deriving the current algebra.}
In a manifestly O$(d,d)$-covariant way, the current is defined as $\bE_I = \eta_{IJ} \partial X^J$ for extended coordinates $X^M = (x^\mu , \tilde{x}_\mu)$. The \textit{current algebra} can be obtained \textit{non-covariantly} by the identification $p_\mu = \partial \tilde{x}_\mu$, as above, and applying the canonical Poisson brackets
\begin{align}
\PB{p_\mu(\sigma)}{x^\nu(\sigma^\prime)} = - \delta_\mu^\nu \delta(\sigma - \sigma^\prime), \quad \PB{p_\mu(\sigma)}{p_\nu(\sigma^\prime)} = \PB{x^\mu(\sigma)}{x^\nu(\sigma^\prime)} =0. \label{eq:STRINGcanonicalPB}
\end{align}
The O$(d,d)$-covariant way to derive the current algebra is as Dirac brackets \cite{Blair:2014kla,Sakatani:2020wah} for the constraints
\begin{equation}
\bPhi_I = \mathbf{P}_I - \bE_I =  \mathbf{P}_I - \eta_{IJ} \partial X^J \label{eq:STRINGconstraint}
\end{equation}
on an \textit{extended} canonical phase space with
\begin{align}
\PB{\mathbf{P}_M(\sigma)}{X^N(\sigma^\prime)} = - c \delta_M^N \delta(\sigma - \sigma^\prime), \quad \PB{\mathbf{P}_M(\sigma)}{\mathbf{P}_N(\sigma^\prime)} = \PB{X^M(\sigma)}{X^N(\sigma^\prime)} =0. \label{eq:STRINGextendedphasespacePB}
\end{align}
The Dirac procedure is necessary, because the constraints $\bPhi_M$ are second class, i.e.
\begin{equation}
\PB{\bPhi_M(\sigma)}{\bPhi_N(\sigma^\prime)} = C_{MN}(\sigma,\sigma^\prime) = - 2c \eta_{MN} \partial \delta(\sigma - \sigma^\prime)
\end{equation}
is invertible. The resulting Dirac brackets for the currents $\bE$ are
\begin{align}
\PB{\bE_M(\sigma)}{\bE_N(\sigma^\prime)}_{D.B.} &= - \int \mathrm{d} \sigma^{\prime\prime} \int \mathrm{d} \sigma^{\prime \prime \prime} \PB{\bE_M(\sigma)}{\bPhi_K(\sigma^{\prime \prime})} (C^{-1})^{KL}(\sigma^{\prime \prime},\sigma^{\prime \prime \prime}) \PB{\bPhi_L(\sigma^{\prime \prime \prime})}{\bE_N(\sigma^\prime)} \nonumber  \\
&= \frac{c}{2} \eta_{MN} \delta \delta(\sigma - \sigma^\prime) \label{eq:STRINGDiracBrackets}. \\
\PB{\bE_M(\sigma)}{X^N(\sigma^\prime)}_{D.B} &= - \frac{c}{2} \delta^N_M \delta(\sigma-\sigma^\prime).
\end{align}
$\PB{\bE_M(\sigma)}{\bE_N(\sigma^\prime)}_{D.B.}$ coincides with \eqref{eq:STRINGCurrentAlgebraDorfman} for $c=2$. As will be important later, the Dirac procedure naturally constructs a Lie bracket (skewsymmetric and satisfying Jacobi identity) but the calculation seems to work only \textit{when neglecting total spatial world-sheet derivative terms}. Hence, for the string one can derive the Dorfman current algebra \eqref{eq:STRINGCurrentAlgebraDorfman} as a Dirac bracket up to such terms.

\paragraph{Generalised flux frame.} A different choice of basis for the currents $\bE_A = {E_A}^I \bE_I$ with ${E_A}^I\in$O$(d,d)$ being a generalised vielbein, i.e. ${E_A}^I {E_B}^J \mathcal{H}_{IJ} = \delta_{AB}$, leads to a diagonalisation of the Hamiltonian $H = \frac{1}{2} \int \mathrm{d} \sigma \delta^{AB} \bE_A(\sigma) \bE_B(\sigma)$ and a twist of the current algebra 
\begin{equation}
\PB{\bE_A(\sigma)}{\bE_B(\sigma^\prime)}_D = \eta_{AB} \partial \delta(\sigma-\sigma^\prime) - {\bF^C}_{AB}(\sigma) \delta(\sigma-\sigma^\prime)
\end{equation}
by the generalised fluxes $\bF_{ABC} = \left(\partial_{[A} {E_B}^I\right) E_{C]I}$ with $\partial_A = {E_A}^M \partial_M = {E_A}^M (\partial_\mu , 0)$. In case of a generalised parallelisable background the $\bF_{ABC}$ can be chosen to be constant. Then, this Hamiltonian formulation takes the form of an  $\mathcal{E}$-model.

\subsection{Summary of results}
All of the above points for the string are generalised to some extend to objects in $E_{d(d)}$, and in part SL$(d+1)$, generalised geometry. A duality invariant Hamiltonian formulation of the $\frac{1}{2}$-BPS $p$-branes is proposed. For that we consider a generalised geometry to some duality group $\mathcal{G}$ on a $d$-dimensional manifold $M$ with coordinates $x^\mu$, including:
\begin{itemize}
\item \textit{two representations of $\mathcal{G}$}: $\mathcal{R}_1$ with indices $K,L,M,...$ and $\mathcal{R}_2$ with indices $\mathcal{K},\mathcal{L},\mathcal{M},...$. The $\mathcal{R}_1$-representation is the one of the generalised tangent bundle, e.g. $(T\oplus \bigwedge^2 T^\star)M$ for the (M-theory section) SL$(5)$-theory, and of the extended coordinates $X^M = (x^\mu,...)$ that are used in exceptional field theory. The representations to the duality groups $\mathcal{G}$, relevant in that paper, are collected in table \ref{table:Representations}.
\begin{table}
\begin{align*}
\begin{array}{c|c|cccc|c}
\mathcal{G} & \text{O}(d,d) & \text{SL}(3)\times \text{SL}(2)& \text{SL}(5) & \text{SO}(5,5) & E_{6(6)} & \text{SL}(p+3) \\ \hline
\text{dim}(M) & d & 3 & 4 & 5 & 6 & p+2 \\ \hline
\text{object} & \text{string} & \multicolumn{4}{c|}{\text{miscellaneous}} & \text{$p$-brane} \\ \hline
\mathcal{R}_1 & \mathbf{2d} & \mathbf{(\bar{3},2)}& \mathbf{10} & \mathbf{16} & \mathbf{27} & \mathbf{\frac{(p+3)(p+2)(p+1)}{3!}} \\
\mathcal{R}_2 & \mathbf{1} & \mathbf{(3,1)}& \mathbf{\bar{5}} & \mathbf{10} & \mathbf{\overline{27}} & \mathbf{\frac{(p+4)(p+3)(p+2)(p+1)}{4!}} \\ \hline
\mathcal{R}_3 & & \mathbf{(1,2)} & \mathbf{5} & \mathbf{\overline{16}} & \mathbf{78} &  \\ 
\mathcal{R}_4 & & \mathbf{(\bar{3},1)}& \mathbf{\overline{10}} & \mathbf{45} & \mathbf{351} &  \\
\vdots & & \vdots & \vdots & \vdots & \vdots & 
\end{array}
\end{align*}
\caption{The relevant representations of duality/generalised geometry groups $\mathcal{G}$ including the begin of the tensor hierachy for the $E_{d(d)}$ groups.}
\label{table:Representations}
\end{table}

\item \textit{$\eta$-symbols $\eta_{\mathcal{M},KL}$ and $\eta^{\mathcal{M},KL}$ -- invariants of $\mathcal{G}$:} From these one can derive the $Y$-tensor that defines the \textit{generalised Lie derivative}, ${Y^{KL}}_{MN} = \eta^{\mathcal{P},KL} \eta_{\mathcal{P},MN}$, and the section condition $\eta^{\mathcal{M},KL} \partial_K f \partial_L g = 0$ on functions  $f,g$ of the extended coordinates $X^M$. 

\item \textit{$\omega$-symbols}: defining a para-Hermitian generalised geometry (dicussed in section \ref{chap:ParaHermitian}) and characterising the difference of the \textit{standard} Lie derivate on the generalised tangent bundle to the \textit{generalised} Lie derivate. Explicite expressions for $\eta$- and $\omega$-symbols (and their decomposition in M-theory and type IIb sections) in the conventions of this paper can be found in appendix \ref{chap:AppendixEdd}.

\item \textit{a generalised metric $\mathcal{H}$}
\end{itemize}
The fundamental objects of such a $p$-brane world-volume theory are currents $\bZ \in \mathcal{R}_1$ that are spatial top-forms on the world-volumes, i.e. (spatial) world-volume $p$-forms:
\begin{align}
\bZ_K (\sigma) &=  \frac{1}{p} \eta_{\mathcal{N},KL} \mathcal{Q}^{\mathcal{N}} \wedge \mathrm{d}X^L. \label{eq:IntroZgeneral}
\end{align}
$\sigma=(\sigma^1, ... , \sigma^p)$ always denote the spatial coordinates of the $p$-brane world-volume, $\int$ the integral of the $p$-dimensional spatial part of the world-volume and $\delta(\sigma-\sigma^\prime)$ the $\delta$-distribution in $p$-dimension.

In terms of this current the proposed Hamiltonian and (reducible) set of spatial diffeomorphism constraints are:
\begin{align}
H &= \frac{1}{2} \int \bZ_K \wedge \star \bZ_L \mathcal{H}^{KL}, \qquad 0 = \eta^{\mathcal{M}, K L} \bZ_K \wedge \star \bZ_L. \label{eq:IntroHamiltonian}
\end{align}
The current $\bZ$ is characterised by a $(p-1)$-form charge $\mathcal{Q} \in \mathcal{R}_2$ and is supposed to satisfy the fundamental current algebra
\begin{align}
\PB{\bZ_M(\sigma)}{\bZ_N(\sigma^\prime)}_D &= \bfeta_{MN} \wedge \mathrm{d} \delta(\sigma-\sigma^\prime) =  \eta_{\mathcal{L},MN} \mathcal{Q}^\mathcal{L} \wedge \mathrm{d}\delta(\sigma - \sigma^\prime) \label{eq:IntroCurrentAlgebraDorfmanGeneral}
\end{align}
such that the model exhibites $\mathcal{G}$-invariance. Indeed, when applying this to functionals \linebreak $\phi = - \int \phi^K \bZ_K$ one reproduces the generalised Lie derivative of $\mathcal{G}$-generalised geometry, if the charge fulfils the condition
\begin{equation}
\mathcal{Q}^{\mathcal{M}} \wedge \mathrm{d}X^N \partial_N = \frac{1}{p} \eta^{\mathcal{M},NP} \eta_{\mathcal{K},LP} \mathcal{Q}^{\mathcal{K}} \wedge \mathrm{d}X^L \partial_N. \label{eq:IntroChargeCond}
\end{equation}
Alternatively, similar to the string, one could also consider the current algebra
\begin{align}
\PB{\bZ_M(\sigma)}{\bZ_N(\sigma^\prime)}_L &= \mathcal{Q}^{\mathcal{L}} \wedge \left( \left( \eta_{\mathcal{L},MN} + \omega_{\mathcal{L},(MN)} \right) \frac{1}{2}\left(\mathrm{d}-\mathrm{d}^\prime\right)  + \omega_{\mathcal{L},[MN]} \frac{1}{2} \left(\mathrm{d}+\mathrm{d}^\prime\right)\right) \delta(\sigma-\sigma^\prime) \label{eq:IntroCurrentAlgebraLieGeneral}
\end{align}
that is a Lie bracket and corresponds to the \textit{standard} Lie derivative on the generalised tangent bundle. For $E_{d(d)}$ with $d\geq 5$ the $\omega$-symbols are not skewsymmetric anymore and, hence, the standard and generalised Lie derivatives on the generalised tangent bundle are {not} equivalent up to total derivative terms anymore.

As a result, it will turn out that current representation for the 'Lie' current algebra \eqref{eq:IntroCurrentAlgebraLieGeneral} generically takes a form like:
\begin{equation}
\bZ_M= \left( p_\mu \  , \ \mathrm{d} x^{\mu_1} \wedge ... \wedge \mathrm{d} x^{\mu_p} \right) \label{eq:IntroCanCurrent}
\end{equation}
from which one can even derive \eqref{eq:IntroCurrentAlgebraLieGeneral} by means of the canonical Poisson brackets for $p$ and $x$. In case of the $p$-brane in the SL$(p+3)$ theory this will be equivalent up to total world-sheet derivative terms to \eqref{eq:IntroCurrentAlgebraDorfmanGeneral}. Then, one can even generalise a derivation by Dirac brackets, as in the string case. This is demonstrated in section \ref{chap:pForm}

For \textit{exceptional currents}, in which we are really interested, such a derivation is \textit{not} possible -- \eqref{eq:IntroCanCurrent} does not form an $E_{d(d)}$-representation in general -- and the currents itself cannot be written in terms of canonical momenta $p(\sigma)$ and field $x(\sigma)$ alone. Solutions to the charge condition \eqref{eq:IntroChargeCond} dictate, that current algebra representations of exceptional symmetries must contain differentials of \textit{'non-geometric'} coordinates, i.e. those that are not part of a section. These could be identified with momentum d.o.f.s of other objects in the theory -- for example for the M5-brane, the charge $\mathcal{Q}$ and the current $\bZ$ seem to contain a combination of geometric ($x^\mu$) and 'non-geometric' ($\tilde{x}_{\mu\nu}$) coordinates, latter are associated to the M2 momentum, i.e. schematically  
\begin{equation}
\bZ_M \sim \left( p_\mu \ , \  \mathrm{d}x^{\mu_1} \wedge \mathrm{d}x^{\mu_2} \wedge \mathrm{d}x^{\mu_3} \wedge \mathrm{d}x^{\mu_4} \wedge \mathrm{d} \tilde{x}_{\mu_3 \mu_4}  \ , \ \mathrm{d}x^{\mu_1} \wedge ... \wedge \mathrm{d}  x^{\mu_5} \right).
\end{equation}
Alternatively, the non-geometric coordinates could be identified with gauge fields associated to M2-branes ending on the M5-brane. The solutions are presented also for all the other $\frac{1}{2}$-BPS $p$-pranes in $d\leq 6$ in detail in section \ref{chap:ExCurrent}. 

For the membrane in SL$(5)$, all of this is shown explicitly in section \ref{chap:SL5}, revisiting partly known results in the literature. It is observed that the key point of the string generalise: derivation of the current algebra from the canonical Poisson brackets and the $\mathcal{E}$-model like form of the current algebra in the generalised flux frame (section \ref{chap:GenFluxes}). The introduction of charges in section \ref{chap:Charges101} achieves manifest SL$(5)$-invariance of the current algebra and together with the known Hamiltonian in that case \cite{Hatsuda:2012vm} motivates the form of the Hamiltonian \eqref{eq:IntroHamiltonian} and the current algebra \eqref{eq:IntroCurrentAlgebraDorfmanGeneral} that is conjectured here to hold more generally.

\section{Membranes in the SL$(5)$-theory}
\label{chap:SL5}
Let us consider the bosonic part of the standard Polyakov-type action of an M2-brane
\begin{align}
S = T \int_\Sigma \left( \frac{1}{2} \mathrm{d}x^\mu \wedge \star  \mathrm{d}x^\nu G_{\mu \nu}(x) + \frac{1}{3!} \mathrm{d}x^\mu \wedge \mathrm{d}x^\nu \wedge \mathrm{d}x^\rho C_{\mu \nu \rho}(x) + \frac{1}{2} \star 1 \right)
\end{align}
 with tension $T$ and coupling to the space-time metric $G$ and a three-form gauge potential $C$. For the target space, we restrict ourselves to work only on the internal part of some compactification on a four-dimensional manifold, so that $\kappa,\lambda,\mu,... = 1,...,d=4$ and work on a three-dimensional world-volume $\Sigma$ with coordinates $\xi = (\tau,\sigma^\alpha)$ and a dynamical world-volume metric. In the following we integrate out latter \cite{Bergshoeff:1987cm} and work in the Hamiltonian formalism. It was shown that Hamiltonian and spatial diffeomorphism constraints take the following form\footnote{The a priori five diffeomorphism constraints \eqref{DiffConstraintM2} can indeed be reduced to two constraints, but only in that redundant form the SL$(5)$-covariance is manifest.} \cite{Hatsuda:2012vm}:
\begin{align}
H &= \frac{1}{2} \int \bZ_K \wedge \star \bZ_L \mathcal{H}^{KL}(G,C)\label{HamiltonianM2} \\
0 &= \eta^{\mathcal{M}, K L} \bZ_K \wedge \star \bZ_L \sim \epsilon^{\mathcal{M} \mathcal{K} \mathcal{K}^\prime \mathcal{L} \mathcal{L}^\prime} \bZ_{\mathcal{K} \mathcal{K}^\prime} \wedge \star \bZ_{\mathcal{L} \mathcal{L}^\prime}. \label{DiffConstraintM2}
\end{align}
Let us explain the notation. From here on all the differential geometric expressions, like $\wedge$, $\star$, '$\mathrm{d}$' or integrals, are to be understood on the (two-dimensional) spatial part of the membrane world-volume.\footnote{The original world-volume metric is integrated out already. A $\star$-operator w.r.t. to a flat metric on the spatial world-sheet is introduced here to simplify the notation. Alternatively $\bZ_K \wedge \star \bZ_L = \bZ_K \bar{\bZ}_L = \bar{\bZ}_K \bZ_L$, where $\bar{\bZ}$ denotes the scalar dual to the top-form $\bZ$.} The indices $\mathcal{K},\mathcal{L},\mathcal{M},...=1,...,5$ denote (fundamental) $\mathcal{R}_2$-indices and $K,L,M,...$ $\mathcal{R}_1 = \mathbf{10}$-indices of SL$(5)$, with decompositions into the four-dimensional \linebreak M-theory target space indices $\kappa,\lambda,\mu,... = 1,...,4$ as $V^M = \frac{1}{\sqrt{2}} V^{\mathcal{M} \mathcal{M}^\prime} = \left(v^\mu , \frac{1}{\sqrt{2}} v_{\mu\mu^\prime} \right)$ and $V^\mathcal{M} = (v^\mu , v^5)$.\footnote{The general conventions for $E_{d(d)}$ $\eta$-symbols are collected in appendix \ref{chap:AppendixEdd}.} $\mathcal{H}^{KL}$ is the (inverse) generalised metric of SL$(5)$ generalised geometry first introduced in \cite{Duff:1990hn}, parameterised for example as
\begin{equation}
\mathcal{H}^{KL}(G,C) =  G^{-\frac{1}{3}} \left( \begin{array}{cc} G^{\kappa \lambda} & - G^{\kappa \mu} C_{\mu \lambda \lambda^\prime} \\ -C_{\kappa \kappa^\prime \mu} G^{\mu \lambda} & 2 G_{\kappa [\lambda} G_{\lambda^\prime] \kappa^\prime} +  C_{\kappa \kappa^\prime \mu} G^{\mu \mu^\prime} C_{\mu^\prime \lambda \lambda^\prime}\end{array} \right).
\end{equation}
The $\eta$-symbols $\eta_{\mathcal{M},KL}$ $\left(\eta^{\mathcal{M},KL}\right)$ are nothing else than the invariant tensor of SL$(5)$ $\eta^{\mathcal{M},\mathcal{K}\mathcal{K}^\prime \mathcal{L}\mathcal{L}^\prime} =  \epsilon^{\mathcal{M}\mathcal{K}\mathcal{K}^\prime \mathcal{L}\mathcal{L}^\prime}$, and can be used to define the $Y$-tensor and the section condition:
\begin{equation}
{Y^{KL}}_{MN} = \eta^{\mathcal{P},KL} \eta_{\mathcal{P},MN} \qquad \eta^{\mathcal{M},KL} \partial_K \otimes \partial_L = 0.
\end{equation}
The \textit{currents} $\bZ \in \mathcal{R}_1$ are a collection of the canonical momenta $p(\sigma)$ and differentials of the fields $x(\sigma)$:
\begin{equation}
\bZ_K = \frac{1}{\sqrt{2} }\bZ_{\mathcal{K} \mathcal{K}^\prime} = \left( p_\kappa \star 1 , \frac{1}{\sqrt{2}}\mathrm{d} x^\kappa \wedge \mathrm{d} x^{\kappa^\prime} \right) \label{eq:SL(5)CurrentRep}
\end{equation}
in the form of spatial 2-forms on the M2-brane world-volume.

\subsection{The SL$(5)$ generalised Lie derivative on the membrane phase space}
We revisit the study in \cite{Hatsuda:2012vm} of the Poisson structure of the above $\bZ_M$ in the following, in order to clarify the appearance of the generalised derivative in the sense of \cite{Osten:2019ayq} by neglecting a topological contribution. The aim is to see how the generalised Lie derivative
\begin{equation}
[\phi_1 , \phi_2 ]^K_D = \phi_{1}^L \partial_L \phi_{2}^K - \phi_{2}^L \partial_L \phi_{1}^K - {Y^{KL}}_{MN} \phi_1^M \partial_L \phi_2^N
\end{equation}
is encoded in the current algebra of an M2-brane. From the canonical (equal time) Poisson brackets of the $p(\sigma)$ and $x(\sigma)$ one derives the current algebra\footnote{using the conventions and identities from appendix \ref{chap:Conventions}}
\begin{align}
\PB{\bZ_K(\sigma)}{\bZ_L(\sigma^\prime)} = \mathrm{d} x^\mu(\sigma) \wedge \left( \frac{1}{2} (\mathrm{d} - \mathrm{d}^\prime) \delta(\sigma - \sigma^\prime) \epsilon_{\mu KL} + \frac{1}{2} (\mathrm{d} + \mathrm{d}^\prime) \delta(\sigma - \sigma^\prime) \omega_{\mu,K L} \right). \label{SL(5)-CurrentAlgebra}
\end{align}
As in the string case we can consider the corresponding 'Dorfman' and 'Courant' brackets:
\begin{align}
\PB{\bZ_K(\sigma)}{\bZ_L(\sigma^\prime)}_D &= \epsilon_{\mu KL} \mathrm{d}x^\mu(\sigma) \wedge \mathrm{d} \delta(\sigma - \sigma^\prime), \label{SL(5)-Dorfman} \\
\PB{\bZ_K(\sigma)}{\bZ_L(\sigma^\prime)}_C &= \epsilon_{\mu KL} \mathrm{d}x^\mu(\sigma) \wedge \frac{1}{2}\left(\mathrm{d}-\mathrm{d}^\prime \right) \delta(\sigma - \sigma^\prime), \label{SL(5)-Courant}
\end{align}
the SL$(5)$-invariant contribution to \eqref{SL(5)-CurrentAlgebra}, when neglecting total spatial derivatives. These names are appropriate because, indeed, they lead to the generalised Lie derivative:
\begin{align}
\PB{\phi_1}{\phi_2}_D &= - 2 \int  \phi_{[1}^L \partial_L \phi_{2]}^K \bZ_K   \epsilon_{\mu KL} + \int \phi_1^K \partial_{\mu^\prime} \phi_2^L \mathrm{d}x^\mu \wedge \mathrm{d}x^{\mu^\prime} \\
\delta^{\mu}_\nu \delta^{\mu^\prime}_{\nu^\prime} \mathrm{d}x^\nu \wedge \mathrm{d}x^{\nu^\prime} \partial_{\mu^\prime} &= \frac{1}{2} {\epsilon^{\mu \mu^\prime}}_{\nu \nu^\prime} \mathrm{d}x^\nu \wedge \mathrm{d}x^{\nu^\prime} \partial_{\mu^\prime} = \frac{1}{4} \epsilon^{\mu mm^\prime nn^\prime} \bZ_{nn^\prime} \partial_{mm^\prime} = \epsilon^{\mu MN} \bZ_N \partial_M \nonumber \\
\Rightarrow \quad \PB{\phi_1}{\phi_2}_D^N (\sigma) &= \phi_{1}^M \partial_M \phi_{2}^N - \phi_{2}^M \partial_M \phi_1^N - {Y^{MN}}_{KL} \phi_1^K \partial_M \phi_2^L
\end{align}
for local functionals $\phi = - \int \phi^K(\sigma) \bZ_K(\sigma)$, by use of the section condition $\tilde{\partial}^{\mu\mu^\prime} \phi(X) = 0$ and the canonical Poisson brackets in $\mathcal{R}_1$-indices, $\PB{\bZ_K(\sigma)}{X^L(\sigma^\prime)} = - \delta^L_K \delta(\sigma - \sigma^\prime)$. 

The second term in \eqref{SL(5)-CurrentAlgebra}, including
\begin{equation}
\omega_{\mu,KL} = \frac{1}{2} \left( \begin{array}{cc} 0 & - {\epsilon_{\mu \kappa}}^{\lambda \lambda^\prime} \\ {\epsilon_{\mu \lambda}}^{\kappa \kappa^\prime} & 0 \end{array} \right),
\end{equation}
breakes the SL$(5)$-invariance of \eqref{SL(5)-CurrentAlgebra}, but makes it a Lie bracket. In fact it will be, as in the O$(d,d)$ case, simply the canonical Lie bracket on $(T\oplus \bigwedge^2 T^\star)M$. To  summarise
\begin{equation}
\PB{\phi_1}{\phi_2}_D = [v_1 , v_2] + \mathcal{L}_{v_1} \xi_2 - \mathrm{d} (\iota_{v_2} \xi_1), \qquad \PB{\phi_1}{\phi_2}_L = [v_1 , v_2] + \mathcal{L}_{v_1} \xi_2 - \mathcal{L}_{v_2} \xi_1
\end{equation}
with $\phi_i = v_i + \xi_i = - \int \phi_i^K \bZ_K \in (T\oplus \bigwedge^2 T^\star)M$. Again, as in the string case, the $\omega$-term corresponds to a boundary/topological contribution, i.e. treated as a distribution 
\begin{equation}
\omega_{\mu,KL} \int \mathrm{d}x^\mu \wedge \mathrm{d} \varphi,
\end{equation}
giving rise to wrapping contributions $\sim \int \mathrm{d}x^\mu \wedge \mathrm{d}x^{\mu^\prime}$. So, the difference between Courant and Dorfman bracket is of this form of a total differential under a spatial world volume integral. This result is completely analogous to the string resp. O$(d,d)$ case \cite{Osten:2019ayq}. There the winding contribution was equivalent to a topological term in the action, necessary for its O$(d,d)$-invariance. The will be discussed further, also for other generalised geometries, in section \ref{chap:ParaHermitian}.

\subsection{Charges and SL(5)-covariance} \label{chap:Charges101}
One can write the current algebra in a manifestly SL$(5)$-invariant way with the use of a charge $\bq_\mathcal{M} \in \overline{\mathcal{R}}_2$. E.g. the \textit{Dorfman} current bracket can be written as
\begin{equation}
\PB{\bZ_K(\sigma)}{\bZ_L(\sigma^\prime)}_D = \epsilon_{\mathcal{M}KL} \bq_{\mathcal{M}^\prime} \mathrm{d} X^{\mathcal{M} \mathcal{M}^\prime}(\sigma) \wedge \mathrm{d} \delta(\sigma - \sigma^\prime)
\end{equation}
or with help of a 1-form valued (SL$(5)$-invariant) 'metric', that can be used to lower the indices,
\begin{equation}
\bm{\eta}_{KL} =  \epsilon_{\mathcal{M}KL} \bq_{\mathcal{M}^\prime} \mathrm{d}X^{\mathcal{M}\mathcal{M}^\prime}, \qquad \text{s.t.} \quad \mathrm{d}\bm{\eta} = 0,
\end{equation}
as
\begin{equation}
\PB{\bZ_K(\sigma)}{\bZ_L(\sigma^\prime)}_D =  \bm{\eta}_{KL} \wedge \mathrm{d} \delta(\sigma - \sigma^\prime).
\end{equation}
Acting on local functionals $\phi = - \int \phi^K \bZ_K$ this gives
\begin{equation}
\PB{\phi_1}{\phi_2}_D = - \int \left( 2 \phi_{[1}^J \partial_J \phi_{2]}^I \  \bZ_I  - \epsilon_{\mathcal{M}KL} \bq_{\mathcal{M}^\prime} \phi_1^K \partial_N \phi_2^L \ \mathrm{d}X^{\mathcal{M}\mathcal{M}^\prime} \wedge \mathrm{d}X^N \right).
\end{equation}
The generalised Lie derivative is reproduced if we identify
\begin{equation}
\mathrm{d}X^{\mathcal{M}\mathcal{M}^\prime} \wedge \mathrm{d}X^{\mathcal{N}\mathcal{N}^\prime} \bq_{\mathcal{M}^\prime} \partial_{\mathcal{N}\mathcal{N}^\prime} = \frac{3}{2} \mathrm{d}X^{[\mathcal{M}|\mathcal{M}^\prime} \wedge \mathrm{d}X^{|\mathcal{N} \mathcal{N}^\prime]} \bq_{\mathcal{M}^\prime} \partial_{\mathcal{N} \mathcal{N}^\prime} = \epsilon^{\mathcal{M}\mathcal{N}\mathcal{N}^\prime K} \bZ_{K} \partial_{\mathcal{N}\mathcal{N}^\prime}, \label{eq:IdentificationM2Currents}
\end{equation}
with the consistency condition $\bq_{[\mathcal{M}} \partial_{\mathcal{N}\mathcal{N}^\prime]}=0$, that was used in the first step. In that way the \textit{charge} $\bq_\mathcal{M}$ was already introduced in \cite{Arvanitakis:2018hfn}.

In the string case we had that the currents $\bE_I = (p_i,\partial x^i)$ are  related to the doubled coordinates like $\bE_I = \eta_{IJ} \partial X^J$. This allowed to derive the (Dorfman, Courant, Lie) brackets of the doubled coordinates $X_I(\sigma)$ and study non-geometric on the string phase space \cite{Blair:2013noa,Blair:2014kla,Osten:2019ayq}. In a similar fashion the 'extended coordinate fields' $X^M$ of SL$(5)$ exceptional field theory are related to the membrane currents $\bZ_M$ by lowering the index with $\bm{\eta}_{MN}$ and taking the spatial world-sheet differential, leading to the following objects and their M-theory decompositions 
\begin{align}
\text{$0$-forms (coordinates):} \qquad X^M & = \left( x^\mu,\frac{1}{\sqrt{2}} \tilde{x}_{\mu \mu^\prime} \right) \nonumber\\
\text{$1$-forms:} \qquad X_M &= \bm{\eta}_{MN} X^N = \epsilon_{\mathcal{L}MN} \bq_{\mathcal{L}^\prime} X^{\mathcal{L}\mathcal{L}^\prime} \mathrm{d}X^N, \qquad \mathbf{Z}^M \equiv \frac{1}{2}\mathrm{d}X^M  \nonumber  \\
\text{$2$-forms (currents):} \qquad \bZ_M &= \frac{1}{2} \bm{\eta}_{MN} \wedge \mathrm{d}X^N = \frac{1}{2} \epsilon_{\mathcal{K}LM} \bq_{\mathcal{K}^\prime} \mathrm{d}X^{\mathcal{K}\mathcal{K}^\prime} \wedge \mathrm{d}X^L \label{eq:M2Currents} \\
& =  \left( \frac{1}{2 }\mathrm{d}\tilde{x}_{\mu \nu} \wedge \mathrm{d} x^{\nu} , \frac{1}{\sqrt{2}} \mathrm{d}x^\mu \wedge \mathrm{d}x^{\mu^\prime} \right), \nonumber
\end{align}
when using the choice of charge $\mathbf{q}_5 = 1$ (with all other components vanishing) that corresponds to the choice of M-theory section used before.\footnote{The last identification of $\bZ_M$ with $\mathrm{d}X^K\wedge \mathrm{d}X^L$ is equivalent to \eqref{eq:IdentificationM2Currents}:
\begin{align*}
\epsilon^{\mathcal{N}\mathcal{R}\mathcal{R}^\prime \mathcal{K}\mathcal{K}^\prime} \bZ_{\mathcal{K}\mathcal{K}^\prime} &= \frac{1}{4} \epsilon^{\mathcal{N}\mathcal{R}\mathcal{R}^\prime \mathcal{K}\mathcal{K}^\prime} \epsilon_{\mathcal{M}\mathcal{L}\mathcal{L}^\prime \mathcal{K}\mathcal{K}^\prime} \bq_{\mathcal{M}^\prime} \mathrm{d} X^{\mathcal{M}\mathcal{M}^\prime} \wedge \mathrm{d} X^{\mathcal{L}\mathcal{L}^\prime} \\
&= \frac{1}{4}\ 2! \ 3! \ \delta^{\mathcal{N}}_{[\mathcal{M}} \delta^{\mathcal{R}}_{\mathcal{L}} \delta^{\mathcal{R}^\prime}_{\mathcal{L}^\prime]} \ \bq_{\mathcal{M}^\prime} \mathrm{d} X^{\mathcal{M}\mathcal{M}^\prime} \wedge \mathrm{d} X^{\mathcal{L}\mathcal{L}^\prime}= 3 \ \bq_{\mathcal{M}^\prime} \mathrm{d} X^{[\mathcal{N}|\mathcal{M}^\prime} \wedge \mathrm{d} X^{|\mathcal{R}\mathcal{R}^\prime]} \left( = 2 \ \epsilon^{\mathcal{N}\mathcal{R}\mathcal{R}^\prime K} \bZ_{K} \right)
\end{align*}
and then as above in \eqref{eq:IdentificationM2Currents}. For that reason the extra factor $\frac{1}{2}$ was needed in the definition of $\bZ$ in \eqref{eq:M2Currents}.}

\subsection{Double reduction of membrane current algebra}
\label{chap:reductionIIa}
As a consistency check for the logic behind identifying the current algebra with objects of exceptional generalised geometry and exceptional field theory, we study the reduction of the membrane to the type IIa string. For that, we perform the usual double dimensional reduction \cite{Duff:1987bx}:
\begin{equation}
x^4(\sigma^1,\sigma^2) = \sigma^2, \quad x^{\underline{\mu}}(\sigma^1 , \sigma^2) = x^{\underline{\mu}}(\sigma^1) \equiv x^{\underline{\mu}}(\sigma).
\end{equation}
for $\underline{\kappa},\underline{\lambda},...=1,2,3$. The membrane current \eqref{eq:SL(5)CurrentRep} becomes
\begin{equation}
\bZ_{\mathcal{M}\mathcal{M}^\prime} = \left( \mathrm{d}x^{\underline{\mu}}\wedge\mathrm{d}x^{\underline{\mu}^\prime} \ , \ \mathrm{d}x^{\underline{\mu}}\wedge\mathrm{d}x^4  \ , \ p_{\underline{\mu}} \ , \ p_4 \right) = \left( 0 \ , \ \mathrm{d}x^{\underline{\mu}}\wedge\mathrm{d} \sigma^2  \ , \ p_{\underline{\mu}} \ , \ 0 \text{\comment{(?)}} \right)
\end{equation}
such that with $\bZ_{\mathcal{M}\mathcal{M}^\prime} \rightarrow \bz_{\mathcal{M}\mathcal{M}^\prime}(\sigma^1) \wedge \mathrm{d} \sigma^2$
\begin{equation}
\bz_{\mathcal{M}\mathcal{M}^\prime} = \big( \underbrace{\epsilon^{\underline{\mu \mu^\prime \nu}} \bz_{\underline{\mu} \underline{\mu}^\prime} \  , \  \bz_{\underline{\mu} 5}}_{\bz_{m}} \ , \ \underbrace{\bz_{\underline{\mu} 4}  \ , \ \bz_{45}}_{\bz_{\tilde{m}}} \big) = \left(  \mathrm{d}x^{\underline{\mu}}  \ , \ p_{\underline{\mu}} \ , \ 0 \ , \ 0 \right) \label{eq:SL(5)CurrentRepStringIIA}
\end{equation}
At the same time the membrane current algebra \eqref{SL(5)-CurrentAlgebra} reduces to
\begin{align}
\PB{\bz_K(\sigma)}{\bz_L(\sigma^\prime)}_L &= \eta_{KL} \frac{1}{2} (\mathrm{d}-\mathrm{d}^\prime)\delta(\sigma-\sigma^\prime) + \frac{1}{2} \omega_{KL} (\mathrm{d}+\mathrm{d}^\prime) \delta(\sigma-\sigma^\prime) \label{eq:SL(5)StringCurrentNonInv} \\
\PB{\bz_K(\sigma)}{\bz_L(\sigma^\prime)}_D &= \eta_{KL} \mathrm{d}\delta(\sigma-\sigma^\prime), \label{eq:SL(5)StringCurrentNonInvDorfman}
\end{align}
with $\eta_{KL} = \epsilon_{4KL}$ and $\omega_{KL} = \omega_{4,KL}$. The only non-vanishing components are:
\begin{equation}
\eta_{\underline{\mu\mu^\prime\nu} 5} = \eta_{\underline{\nu} 5\underline{\mu\mu^\prime}} = \omega_{\underline{\mu\mu^\prime\nu} 5} = - \omega_{\underline{\nu} 5\underline{\mu\mu^\prime}} = \epsilon_{\underline{\mu \mu^\prime \nu}}
\end{equation}
or in the conventions of \eqref{eq:SL(5)CurrentRepStringIIA}
\begin{align*}
\eta_{KL} &= \left( \begin{array}{cc} \underline{\eta}_{kl} & 0 \\ 0 & 0 \end{array} \right), \quad \underline{\eta} = \left( \begin{array}{cc} 0 & \mathbbm{1}_3 \\ \mathbbm{1}_3 & 0 \end{array} \right); \qquad 
\omega_{KL} &= \left( \begin{array}{cc} \underline{\omega}_{kl} & 0 \\ 0 & 0 \end{array} \right), \quad \underline{\omega} = \left( \begin{array}{cc} 0 & -\mathbbm{1}_3 \\ \mathbbm{1}_3 & 0 \end{array} \right)
\end{align*}
which, restricted to the $k,l,...= ({ }_{\underline{\kappa}},{ }^{\underline{\kappa}}),({}_{\underline{\lambda}},{}^{\underline{\lambda}}),...$ indices, is the canonical O$(3,3)$ metric and the components of the canonical symplectic form $\omega$ as expected from the string discussion. $T$-duality transformations are defined by ${M_M}^N \in$SL$(5)$ with ${M_M}^K {M_N}^L \eta_{KL} = \eta_{MN}$.

\paragraph{String charges and SL$(5)$-covariance.} The type IIa discussion above motivates 
\begin{align}
\PB{\bz_K(\sigma)}{\bz_L(\sigma^\prime)} &= \bq^{\mathcal{M}} \left(\eta_{\mathcal{M},KL} \frac{1}{2} (\mathrm{d}-\mathrm{d}^\prime)\delta(\sigma-\sigma^\prime) + \frac{1}{2} \omega_{\mathcal{M},KL} (\mathrm{d}+\mathrm{d}^\prime) \delta(\sigma-\sigma^\prime) \right) \\
\PB{\bz_K(\sigma)}{\bz_L(\sigma^\prime)}_D &= \bq^{\mathcal{M}} \eta_{\mathcal{M},KL} \mathrm{d}\delta(\sigma-\sigma^\prime), \label{eq:SL(5)DorfmannStringCurrent}
\end{align}
as the SL$(5)$ string current algebra using the string charge \linebreak $\bq^\mathcal{M} \in \mathcal{R}_2$, fulfilling $\bq^\mathcal{M} \partial_{\mathcal{M}\mathcal{M}^\prime} = 0$ \cite{Arvanitakis:2018hfn}. The SL$(5)$-invariant M$\rightarrow$IIa reduction condition would be something like
\begin{equation}
\bq^\mathcal{M} \sigma^2 = \bq_{\mathcal{M}^\prime} X^{\mathcal{M} \mathcal{M}^\prime}.
\end{equation}
Let us proceed in the same way as for the membrane current algebra, and try to reproduce the generalised Lie derivative form the Dorfman current algebra \eqref{eq:SL(5)DorfmannStringCurrent}
\begin{align}
\PB{\phi_1}{\phi_2} &= -\int \left( 2\phi_{[1}^K \partial_K \phi_{2]}^L \bz_L - \bq^\mathcal{P} \eta_{\mathcal{P},KL} \phi_1^K \partial_M \phi_2^L \mathrm{d}X^M \right) \label{eq:StringCurrentGeneralisedLieDer1}
\end{align}
We need to make a similar identification as in the membrane case. Let us define $\eta_{KL} = \frac{1}{2} \bq^p \epsilon_{pKL}$ in an SL$(5)$-invariant way. Then it is natural to define
\begin{equation}
\bz_K = \bm{\eta}_{KL} \mathrm{d}X^L = \eta_{\mathcal{M},KL} \mathbf{q}^\mathcal{M} \mathrm{d}X^L,
\end{equation}
from which 
\begin{align*}
\epsilon^{\mathcal{P}\mathcal{M}\mathcal{M}^\prime \mathcal{N}\mathcal{N}^\prime} \bz_{\mathcal{N}\mathcal{N}^\prime} &= \frac{1}{4} \epsilon^{\mathcal{P}\mathcal{M}\mathcal{M}^\prime \mathcal{N}\mathcal{N}^\prime} \bq^\mathcal{K} \epsilon^{\mathcal{K}\mathcal{L}\mathcal{L}^\prime \mathcal{N}\mathcal{N}^\prime} \mathrm{d}X^{\mathcal{L}\mathcal{L}^\prime} = \frac{1}{4} 3 ! \ 2! \ \bq^{[\mathcal{P}} \mathrm{d} X^{\mathcal{M}\mathcal{M}^\prime ]} = 3 \bq^{[\mathcal{P}} \mathrm{d} X^{\mathcal{M}\mathcal{M}^\prime ]} \\
3 \bq^{[\mathcal{P}} \mathrm{d} X^{\mathcal{M}\mathcal{M}^\prime ]} \partial_{\mathcal{M}\mathcal{M}^\prime} &= \bq^{\mathcal{P}} \mathrm{d} X^{\mathcal{M}\mathcal{M}^\prime} \partial_{\mathcal{M}\mathcal{M}^\prime}
\end{align*} 
follows, making use of the charge condition $\bq^\mathcal{M} \partial_{\mathcal{M}\mathcal{M}^\prime} = 0$. Then, as wished,  \eqref{eq:StringCurrentGeneralisedLieDer1} becomes
\begin{align}
\PB{\phi_1}{\phi_2} &= - \int\left( 2 \phi_{[1}^K \partial_K \phi_{2]}^N  - {Y^{MN}}_{KL} \phi_1^K \partial_M \phi_2^L \right) \bz_N. \label{eq:StringCurrentGeneralised}
\end{align}

\subsection{Twist by generalised vielbein and the embedding tensor} \label{chap:GenFluxes}
In analogy to the string case \cite{Osten:2019ayq}, we aim to diagonalise the Hamiltonian and all the constraints via $\bZ_A(\sigma) = {E_A}^K(\sigma) \bZ_K(\sigma)$ in order to characterise the model via a twist of the current algebra and to bring the model into the form of an $\mathcal{E}$-model. ${E_A}^K$ are generalised frame fields\footnote{The SL$(5)$-vielbeine and their invariance conditions, used here, are:
\begin{align*}
\text{'little' vielbein:}&{}\qquad {E_{\mathcal{A}_1}}^{\mathcal{M}_1}  ... {E_{\mathcal{A}_5}}^{\mathcal{M}_5} \epsilon^{\mathcal{A}_1 ... \mathcal{A}_5} = \epsilon^{\mathcal{M}_1 ... \mathcal{M}_5} \\
\text{'big' vielbein:}&{}\qquad {E_A}^K = \frac{1}{2} {E_{\mathcal{A}\mathcal{A}^\prime}}^{\mathcal{K}\mathcal{K}^\prime} = {E_{[\mathcal{A}}}^\mathcal{K} {E_{\mathcal{A}^\prime]}}^{\mathcal{K}^\prime}, \qquad {E_A}^K {E_B}^L {E_M}^C {E_N}^D {Y^{MN}}_{KL} = {Y^{CD}}_{AB}.
\end{align*}
In a frame, defined by such a vielbein, also the following object, a 'little' current, appears:
\begin{align*}
j^\mathcal{C} &= {E^\mathcal{C}}_\mu \mathrm{d}x^\mu \quad \text{with} \quad \mathrm{d}(\eta_{\mathcal{C},AB} j^\mathcal{C}) = \mathrm{d}(\epsilon_{\mu KL} {E_A}^K {E_B}^L \mathrm{d}x^\mu) = -{Y^{MN}}_{KL} \partial_M ({E_A}^K {E_B}^L) \bZ_N.
\end{align*}
$\mathcal{A},\mathcal{B},...$ are the flat (generalised flux frame) $\mathcal{R}_2$ SL$(5)$-indices and $A = [\mathcal{A}\mathcal{A}^\prime],...$ flat (generalised flux frame) $\mathcal{R}_1$ SL$(5)$-indices.} of the SL$(5)$-theory. For the canonical (Lie) current algebra \eqref{SL(5)-CurrentAlgebra} we have
\begin{align}
\PB{\bZ_A(\sigma)}{\bZ_B(\sigma^\prime)}_L
&= \frac{1}{2} \left( j^\mathcal{C}(\sigma) \wedge  \mathrm{d} \delta(\sigma - \sigma^\prime) - j^\mathcal{C}(\sigma^\prime) \wedge \mathrm{d}^\prime \delta(\sigma - \sigma^\prime) \right) \eta_{\mathcal{C},AB} \label{SL(5)-CurrentAlgebraTwisted} \\
&{}\quad + \mathrm{d}j^\mathcal{C} \wedge \frac{1}{2} (\mathrm{d} + \mathrm{d}^\prime) \delta(\sigma - \sigma^\prime) \omega_{\mathcal{C},AB}(\sigma,\sigma^\prime) \nonumber \\
&{} \quad - {\bF^C}_{[AB]}(\sigma) \bZ_C(\sigma) \delta(\sigma - \sigma^\prime). \nonumber
\end{align}
The $\epsilon$-symbol is SL$(5)$-invariant, whereas $\omega_{\mu,AB}(\sigma,\sigma^\prime) = {E_A}^K(\sigma) {E_B}^L(\sigma^\prime) \omega_{\mu,KL}$ is not SL$(5)$-invariant but again necessary for \eqref{SL(5)-CurrentAlgebraTwisted} to be a Lie bracket. The twist is characterised by the  SL$(5)$ generalised fluxes \cite{Malek:2012pw,Blair:2014zba} $[E_A , E_B]_D = {\bF^C}_{AB} E_C$, i.e.
\begin{equation}
{\bF^C}_{AB} = 2 {E_N}^C \partial_{[A} {E_{B]}}^N - {Y^{CD}}_{AE} {E_N}^E \partial_D {E_B}^N.
\end{equation}
From this definition it is quite obvious to see, why they should appear generically in the current algebra that reproduces the generalised Lie derivative. The twists can also be defined for the Courant algebroid brackets
\begin{align}
\PB{\bZ_A(\sigma)}{\bZ_B(\sigma^\prime)}_D &= j^c(\sigma) \wedge \mathrm{d} \delta(\sigma - \sigma^\prime) \epsilon_{c AB} - {\bF^C}_{AB}(\sigma) \bZ_C(\sigma) \delta(\sigma - \sigma^\prime) \label{SL(5)-DorfmanTwisted} \\
\PB{\bZ_A(\sigma)}{\bZ_B(\sigma^\prime)}_C &= j^c(\sigma) \wedge \frac{1}{2} \left( \mathrm{d} - \mathrm{d}^\prime \right) \delta(\sigma - \sigma^\prime) \epsilon_{c AB} - {\bF^C}_{AB}(\sigma) \bZ_C(\sigma) \delta(\sigma - \sigma^\prime). \label{SL(5)-CourantTwisted}
\end{align}
Some kind of Courant algebroid conditions will put conditions on the ${\mathbf{F}^C}_{AB}$, corresponding to a (dynamical) Bianchi identity of these fluxes.

The type of finite-dimensional algebras, as for example the recently discussed exceptional Drinfeld algebras \cite{Sakatani:2019zrs,Malek:2019xrf,Sakatani:2020iad,Bakhmatov:2020kul,Blair:2020ndg,Musaev:2020bwm,Malek:2020hpo,Sakatani:2020wah,Gubarev:2020ydf,Musaev:2020nrt}, are contained in the current algebra as the algebra of the zero modes $z_A = - \int \bZ_A$ in case $\bF$ is \textit{constant}:
\begin{align}
\{ z_A , z_B \}_D &= \int {\bF^C}_{AB} (\sigma) \bZ_C (\sigma) =  {\bF^C}_{AB} z_C \nonumber \\
\{ z_A , z_B \}_C &= {\bF^C}_{[AB]} z_C, \label{eq:DrinfeldAlgebra} \\
\{ z_A , z_B \} &= {\bF^C}_{[AB]} z_C + \int \eta^{\mathcal{E},CD} \partial_D \left(\omega_{\mathcal{E},AB}\right) \bZ_C \nonumber. 
\end{align}
The boundary/winding contribution to the zero mode algebra ensures (similarly to \cite{Osten:2019ayq} for the string case) that this zero mode algebra in the last line is indeed a Lie algebra. The $\{z_A , z_B \}_D$ bracket is the one appearing in the recent discussions on exceptional Drinfeld algebras.

\section{Canonical and exceptional currents}
We aim to apply the same concepts as in section 2 to the world-volume theories of all the $\frac{1}{2}$-BPS objects in string and M-theory, and investigate how their current algebra can be understood to incorporate generalised geometry structures.

There are two generalisations that we will be concerned with. Of course, the key interest lies in an $E_{d(d)}$-invariant description of world-volume theories of objects in type II string and M-theory. However, some features are shared by the simpler case of $p$-branes in SL$(p+3)$ generalised geometry, considered in \cite{Bonelli:2005ti,Lee:2014mla}. In all cases here, the $Y$-tensor can be defined in terms of $\eta$-symbols
\begin{equation}
{Y^{MN}}_{KL} = \eta^{\mathcal{P},MN} \eta_{\mathcal{P},KL}.
\end{equation}
$\mathcal{R}_1$-indices are denoted by $K,L,M...$ and $\mathcal{R}_2$-indices by  $\mathcal{K},\mathcal{L},\mathcal{M},... \ $, see table \ref{table:Representations}. Let us first settle our conventions for the appearing duality groups, here.

\paragraph{SL$(p+3)$ generalised geometry} is the geometry on the generalised tangent bundle \linebreak $\left( T \oplus \bigwedge^p T^\star \right)M$. The $\eta$-symbol and the corresponding $\omega$-symbol, that will be used in the following, are
\begin{align}
\eta_{\kappa_1 ... \kappa_{p-1},MN} = \left( \begin{array}{cc} 0 & \frac{\delta_{\kappa_1 ... \kappa_{p-1} \mu}^{\nu_1 ... \nu_p}}{\sqrt{p!}} \\ \frac{\delta_{\kappa_1 ... \kappa_{p-1} \nu}^{\mu_1 ... \mu_p}}{\sqrt{p!}} & 0 \end{array} \right), \ \quad
\omega_{\kappa_1 ... \kappa_{p-1},MN} = \left( \begin{array}{cc} 0 & - \frac{\delta_{\kappa_1 ... \kappa_{p-1} \mu}^{\nu_1 ... \nu_p}}{\sqrt{p!}} \\ \frac{\delta_{\kappa_1 ... \kappa_{p-1} \nu}^{\mu_1 ... \mu_p}}{\sqrt{p!}} & 0 \end{array} \right), \label{eq:SL(d)Symbols}
\end{align}
where other components are supposed to vanish and the following decompositions of $\mathcal{R}_1$- and $\mathcal{R}_2$-indices into spacetime indices $\kappa,\lambda,\mu=1,...,d=p+2$ are made:
\begin{align*}
X^M &= \left(x^\mu,\frac{\tilde{x}_{\mu_1 ... \mu_p}}{\sqrt{p!}}\right) \quad \in \mathcal{R}_1, \quad \mathcal{Q}^\mathcal{M}= \left( \frac{Q^{\mu_1 ... \mu_{p-1}}}{\sqrt{(p-1)!}} , \frac{Q^{\mu_1 \mu_2 \mu_3 \mu_4}}{\sqrt{4!}} \right) \quad \in \mathcal{R}_2.
\end{align*}
$\mathcal{R}_1$ and $\mathcal{R}_2$ correspond to the three- resp. four-fold skewsymmetric representation of \linebreak SL$(p+3)$. That generalises the $p=2$-case, that is the membrane case from section \ref{chap:SL5}. The generalised Lie derivative $[\phi_1 ,\phi_2]_D$ takes the same form as for O$(d,d)$ and SL$(5)$ for all $p$: 
\begin{equation}
[\phi_1,\phi_2]_D = [v_1 , v_2] + \mathcal{L}_{v_1} \xi_2 - \mathrm{d} (\iota_{v_2} \xi_1) \label{eq:SL(d)GenLieDerivative}
\end{equation}
with $\phi_i = v_i + \xi_i \in \left( T \oplus \bigwedge^p T^\star \right)M$.

\paragraph{$E_{d(d)}$ generalised geometry.} The $\eta$-symbols and conventions for decomposition of the $\mathcal{R}_1$- and $\mathcal{R}_2$-representation $E_{d(d)}$ for $d\leq 6$ in the type IIb and M-theory section are collected in appendix \ref{chap:AppendixEdd}. The representations of the duality groups that will be revelant are collected in table \ref{table:Representations}. Mainly the representations $\mathcal{R}_1$ and $\mathcal{R}_2$ will matter in the following, whereas the higher representations of the tensor hierarchy will not appear in the following. For $d\leq 6$ the generalised Lie derivative always takes the form (in the M-theory section):
\begin{equation}
[\phi_1,\phi_2]_D = [v_1 , v_2] +  \left( \mathcal{L}_{v_1} \omega_2 - \mathrm{d} (\iota_{v_2} \omega_1) \right) + \left( \mathcal{L}_{v_1} \xi_2 - \mathrm{d} (\iota_{v_2} \xi_1) + \omega_1 \wedge \mathrm{d} \omega_2 \right) \label{eq:EddGenLieDerivative}
\end{equation}
with $\phi_i = v_i + \omega_i + \xi_i \in \left( T \oplus \bigwedge^2 T^\star \oplus \bigwedge^5 T^\star \right)M$.

\paragraph{The general setup.} The proposition of this section is a Hamiltonian formulation of any $p$-brane object in a duality covariant way. The obvious conjecture, generalising the membrane in SL$(5)$, for the Hamiltonian and spatial diffeomorphism constraints is 
\begin{align}
H &= \frac{1}{2} \int \bZ_K \wedge \star \bZ_L \mathcal{H}^{KL}, \qquad 0 = \eta^{\mathcal{M}, K L} \bZ_K \wedge \star \bZ_L. \label{eq:Hamiltonian}
\end{align}
for the generalised metric $\mathcal{H}$ and the $\eta$-symbols $\eta_{\mathcal{M},KL}$ of that theory. The currents $\bZ_M$ are spatial world-volume $p$-forms. In the following, these currents $\bZ$ and their algebra will be the main focus having two main questions in mind:
\begin{itemize}
\item How should the currents $\bZ$ and a currents algebra look, such that they form some kind of representation of the exceptional symmetry? I.e. how must the currents and their algebra look, such that bracket of $\phi = - \int \phi^M \bZ_M$ will reproduce generaslied Lie brackets \eqref{eq:SL(d)GenLieDerivative} or \eqref{eq:EddGenLieDerivative}?

\item Can these currents be constructed by means of the canonical Poisson structure? I.e. Will the currents be in generality derivatives of coordinate fields $x(\sigma)$ and their canonical momenta $p(\sigma)$, arranged into representations of the duality group as in \eqref{eq:M2Currents} for the membrane?
\end{itemize}

\subsection{Para-Hermitian generalised geometries}
\label{chap:ParaHermitian}
In order to answer the above questions we first go a step back, and reconsider why the construction of the generalised Lie derivative from the canonical Poisson structure (up to total derivative terms) worked for the string and the membrane.

In these constructions a non duality-invariant $\omega$-symbol appeared, ensuring the Lie bracket properties of the current algebra. Let us investigate how the invariants $\eta_{\mathcal{M},KL}$ and the $\omega$-symbols combine to produce a Lie bracket. This is part of so-called para-Hermitian geometry \cite{Vaisman:2012px,Freidel:2018tkj,Svoboda:2018rci,Marotta:2018myj,Hassler:2019wvn,Marotta:2019eqc}.

\subsubsection{O$(d,d)$}
We define the projector
$${P^{KL}}_{MN} = \frac{1}{2} \left( {Y^{KL}}_{MN} + {\Omega^{KL}}_{MN} \right)$$ 
with ${Y^{KL}}_{MN} = \eta^{KL} \eta_{MN}$ and ${\Omega^{KL}}_{MN} = \eta^{KL} \omega_{MN}$ for O$(d,d)$ generalised geometry. This notation is chosen in a way that it directly generalises to exceptional generalised geometry. The projector has the following properties:
$${P^{KL}}_{MN} = {P^{LK}}_{MN}, \quad {P^{KL}}_{MN} \partial_K \otimes \partial_L = 0 \quad \text{(section condition)}$$
Using this section condition, one can derive the identities:
\begin{align}
{P^{KL}}_{MN} {P^{NP}}_{RS} \partial_P &= {P^{KL}}_{RS} \left(\begin{array}{c} 0 \\\tilde{\partial}^m \end{array} \right) \approx 0, \label{eq:OddProjId1} \\
{P^{KL}}_{NM} {P^{NP}}_{RS} \partial_P &= {P^{KL}}_{RS} \left(\begin{array}{c} \partial_m \\ 0 \end{array} \right) \approx {P^{KL}}_{RS} \partial_M. \label{eq:OddProjId2}
\end{align}
In comparison to the standard Courant algebroid structure, there are now two total derivative\footnote{Contracting these with the O$(d,d)$-currents $\bE$ makes it obvious that they correspond to total derivatives under the spatial world-volume integrals:
\begin{align*}
\llrrparen{\phi_1,\phi_2}  &= \int \llrrparen{\phi_1,\phi_2}^K \bE_K =   2 \eta_{ MN} \int \mathrm{d} \left(\phi_1^M \phi_2^N\right) = 2 \int \mathrm{d} \left( \phi_1 \bullet \phi_2 \right) \\
\llrrbracket{\phi_1,\phi_2} &= \int \llrrbracket{\phi_1,\phi_2}^K \bE_K =   2 \omega_{MN} \int \mathrm{d} \left(\phi_1^M \phi_2^N\right) = 2 \int \mathrm{d} \left( \phi_1 \circ \phi_2 \right). \nonumber
\end{align*}} bilinear objects:
\begin{align}
\llrrparen{\phi_1,\phi_2}^K &= \frac{1}{2} {Y^{KL}}_{MN} \partial_L\left (\phi_1^M \phi_2^N\right), \qquad  \llrrbracket{\phi_1,\phi_2}^K = \frac{1}{2} {\Omega^{KL}}_{MN} \partial_L\left (\phi_1^M \phi_2^N\right) \label{eq:TopolTermsOdd}
\end{align}
Besides the standard Dorfman bracket $[\phi_1,\phi_2]^K_D$, which is defined in terms of the $Y$-tensor, one can define a bracket in terms of the projector $P$:
\begin{align}
[\phi_1,\phi_2]^K_L &= \phi_{1}^L \partial_L \phi_{2}^K - \phi_{1}^L \partial_L \phi_{2}^K + 2 {P^{KL}}_{MN} \phi_{[1}^M \partial_L \phi_{2]}^N
\end{align}
By the identities \eqref{eq:OddProjId1} and \eqref{eq:OddProjId2} this is indeed a Lie bracket and in fact
\begin{equation}
[\phi_1,\phi_2]_L = [v_1 , v_2] +  \mathcal{L}_{v_1} \xi_2 - \mathcal{L}_{v_2} \xi_1 = [\phi_{[1},\phi_{2]}]_D +\llrrbracket{\phi_1,\phi_2} \equiv [\phi_{1},\phi_{2}]_C +\llrrbracket{\phi_1,\phi_2}, \label{eq:OddLieDerivative}
\end{equation}
the standard Lie derivative on $\left( T \oplus T^\star \right)M$ with $\phi_i = v_i + \xi_i \in \left( T \oplus T^\star \right)M$. Also, $[\phi_1,\phi_2]_L \sim [\phi_1,\phi_2]_C$ up to total derivative terms in the O$(d,d)$ case.

To summarise: the para-Hermitian geometry connected to the pair $(\eta,\omega)$ defines a (standard) Lie algebroid over $TM \oplus T^\star M$. Connected to that, Lie bracket by total derivative terms, are other algebraic entities with interesting properties, including the standard Courant algebroid over $M$.

\paragraph{$\omega$-term and the section.} The crucial point of the '$\omega$-geometry' is that it, in contrast to the standard approach to generalised geometry or double field theory, allows for a reconstruction of the section from the choice of $\omega$ and vice versa.

We choose the ${P^{KL}}_{MN}$ as the fundamental object obeying the identities \eqref{eq:ProjectorId} and start with the standard section $\partial_N \approx \left( \partial_n , 0 \right)$. Then, as ${P^{KL}}_{MN} = \frac{1}{2}\eta^{KL}(\eta_{MN} + \omega_{MN})$, the identities \eqref{eq:ProjectorId} imply
\begin{equation}
\omega = \left( \begin{array}{cc} 2B & -\mathbbm{1} \\ \mathbbm{1} & 0
\end{array}\right)
\end{equation}
for some skewsymmetric matrix $B$. If we took an arbitrary section $\partial^\prime_M = {M_M}^N \partial_N$, $\omega$ transforms as $\omega^\prime = M \cdot \omega \cdot M^T$, for $M \in$ O$(d,d)$. Up to a (constant) $B$-shift, the choice of section determines the form of $\omega$ and vice versa. This $B$-shift symmetry is also a well-known property of a Courant algebroid.

A conceptual consequence is that one can reconstruct the Lie algebroid structure of the current algebra from the standard Courant algebroid \textit{plus} a choice of section.

\paragraph{Para-Hermitian and para-K\"ahler geometries.} The current algebra is characterised by the pair $(\eta,\omega)$ and could be completed to a compatible triple $(\eta,\omega,I)$ by ${I^M}_N = \eta^{MK} \omega_{KN}$. If $I$ is a real structure, $I^2 = \mathbbm{1}$, the geometry is called \textit{para-Hermitian}, if $\mathrm{d}\omega = 0$ \textit{para-K\"ahler} \cite{Vaisman:2012px,Svoboda:2018rci,Hassler:2019wvn}.

In addition, a string model is defined by a generalised metric $\mathcal{H}$ in the Hamiltonian formalism. Recently, Born geometry was introduced as para-K\"ahler geometry of the tripel $(\eta,\omega,\mathcal{H})$, subject to the conditions \cite{Freidel:2018tkj}
\begin{equation}
\eta^{-1} \mathcal{H} = \mathcal{H}^{-1} \eta, \quad \omega^{-1} \mathcal{H} = - \mathcal{H}^{-1} \omega.
\end{equation}
A central result of \cite{Freidel:2018tkj,Svoboda:2018rci} was that, following from this, there exists a frame in which all the defining structures take a canonical form:
\begin{equation}
\mathcal{H} = \mathbbm{1}, \quad \omega = \left(\begin{array}{cc} 0 & -\mathbbm{1} \\ \mathbbm{1} & 0 \end{array} \right), \quad \eta = \left(\begin{array}{cc} 0 & \mathbbm{1} \\ \mathbbm{1} & 0 \end{array} \right).
\end{equation}
The input that we obtain from the Hamiltonian formulation of the string is different, though. In the generalised metric formulation, in which we worked so far, -- meaning canonical coordinates on the phase space and background information encoded in the Hamiltonian via the generalised metric -- we get $\eta$ and $\omega$ in their canonical form and $\mathcal{H}(G,B)$ in a general background dependent form. So, unless we are in flat (or toroidal) space, where we can choose $\mathcal{H} = \mathbbm{1}$, the classical phase space geometry of the string $\sigma$-model is \textit{not} described by Born geometry.

\subsubsection{SL$(p+3)$ and $E_{d(d)}$}
In total analogy to the O$(d,d)$ case, we study the role that the $\omega$-term plays in exceptional generalised geometry. The main question will be whether a similar relation to \eqref{eq:OddLieDerivative} holds, i.e. whether generalised and standard Lie derivatives are the same up to total derivative terms.

We use the definitions of $\eta$- and $\omega$-symbols in \eqref{eq:SL(d)Symbols} and appendix \ref{chap:AppendixEdd} in order to define the projector ${P^{KL}}_{MN} = \frac{1}{2} \left( {Y^{KL}}_{MN} + {\Omega^{KL}}_{MN} \right) = \frac{1}{2} \eta^{\mathcal{P},KL} \left( \eta_{\mathcal{P},MN} +   \omega_{\mathcal{P},MN} \right)$. This projector has the following properties:
$${P^{KL}}_{MN} = {P^{LK}}_{MN}, \quad {P^{KL}}_{MN} \partial_K \otimes \partial_L = 0 \quad \text{(section condition)}$$
Working on a solution to the section condition one derives
\begin{align}
{P^{KL}}_{MN} {P^{NP}}_{RS} \partial_L \otimes \partial_P & \approx 0 \nonumber \\
{P^{KL}}_{NM} {P^{NP}}_{RS} \partial_{(L} \otimes \partial_{P)} &\approx {P^{KL}}_{RS} \partial_{(L} \otimes \partial_{M)} \label{eq:ProjectorId} \\
\text{e.g.} \quad {P^{KL}}_{NM} {P^{NP}}_{RS} \left(\partial_{L} \phi_{[1}^M \right) \left( \partial_{P} \phi_{2]}^S \right) &\approx {P^{KL}}_{RS} \left(\partial_{L} \phi_{[1}^P \right) \left( \partial_{P} \phi_{2]}^S \right)
\end{align}
As is also the case in the usual exceptional generalised geometry in identities regarding the $Y$-tensor \cite{Berman:2012vc}, these identities involing the projector $P$ are a weaker than in the string case. 

Instead of taking a given section and computing these identities, one could again take the opposite route and choose the ${P^{KL}}_{MN}$ as the fundamental object obeying the identities \eqref{eq:ProjectorId}. Then one sees that a choice of $\omega$ is equivalent to a choice of (M-theory or IIb) section up to a gauge-transformation of the three-form gauge fields.

Again, one can define a total derivative object 
\begin{align}
\llrrbracket{\phi_1,\phi_2}^K &= \frac{1}{2} {\Omega^{KL}}_{MN} \partial_L\left (\phi_1^M \phi_2^N\right) \label{eq:TopolTerms}
\end{align}
and, besides the standard Dorfman brackets \eqref{eq:SL(d)GenLieDerivative} and \eqref{eq:EddGenLieDerivative}, a bracket in terms of the projector $P$:
\begin{align}
[\phi_1,\phi_2]^K_L &= \phi_{1}^L \partial_L \phi_{2}^K - \phi_{1}^L \partial_L \phi_{2}^K + 2 {P^{KL}}_{MN} \phi_{[1}^M \partial_L \phi_{2]}^N.
\end{align}
By the identities \eqref{eq:ProjectorId} this is a Lie bracket. From here on, the stories slightly diverge:
\begin{itemize}
\item For the SL$(p+3)$-theory, we have 
\begin{equation}
[\phi_1,\phi_2]_L = [v_1 , v_2] +  \mathcal{L}_{v_1} \xi_2 - \mathcal{L}_{v_2} \xi_1 = [\phi_{1},\phi_{2}]_C +\llrrbracket{\phi_1,\phi_2}, \label{eq:SLdLieDerivative}
\end{equation}
with $\phi_i = v_i + \xi_i \in \left( T \oplus \bigwedge^p T^\star\right)M$ as in the O$(d,d)$-case.

\item For the \textbf{$E_{d(d)}$}-theory, on the other hand, we have
\begin{equation}
[\phi_1,\phi_2]_L = [v_1 , v_2] + (\mathcal{L}_{v_1} \omega_2 - \mathcal{L}_{v_2} \omega_1) +  (\mathcal{L}_{v_1} \xi_2 - \mathcal{L}_{v_2} \xi_1)  \label{eq:EddLieDerivative}
\end{equation}
with $\phi_i = v_i + \omega_i + \xi_i \in \left( T \oplus \bigwedge^2 T^\star \oplus \bigwedge^5 T^\star\right)M$. So, essentially, the para-Hermitian $E_{d(d)}$-geometry is basically \textit{the same} as multiple copies of para-Hermitian $SL(p+3)$-geometry put together. But, the key difference is that the generalised Lie derivative of $E_{d(d)}$-generalised geometry and the standard Lie derivative on the extended tangent bundle do \textbf{not} only differ by a total derivative term: $[\phi_1,\phi_2]_L \neq [\phi_{1},\phi_{2}]_C +\llrrbracket{\phi_1,\phi_2}$. The reason is, that the 'interaction' term $\omega_{[1} \wedge \mathrm{d} \omega_{2]}$ in (the Courant bracket version of) \eqref{eq:EddGenLieDerivative} is \textit{not} a total derivative. On the level of the $\omega$-symbol this has also the consequence that the $\omega$-symbols are in general \textit{not} skewsymmetric: $\omega_{\mathcal{M},KL} \neq - \omega_{\mathcal{M},LK}$, see appendix \ref{chap:AppendixEdd} for the explicite expressions.
\end{itemize}

\subsection{Charges, currents and the (generalised) Lie derivatives}
As currents of $p$-branes we postulate, motivated by our investigation of the M2-brane and the type IIa and type IIb string in the SL$(5)$-theory in section \ref{chap:SL5}, spatial top-forms on the world-volumes, i.e. (spatial) world-volume $p$-forms:
\begin{align}
\bZ_K (\sigma) &=  \frac{1}{p} \eta_{\mathcal{N},KL} \mathcal{Q}^{\mathcal{N}} \wedge \mathrm{d}X^L. \label{eq:DefZgeneral}
\end{align}
For sake of simplicity of notation, $\sigma=(\sigma^1, ... , \sigma^p)$ always denote the spatial coordinates of the $p$-brane world-volume, $\int$ the integral of the $p$-dimensional spatial part of the world-volume and $\delta(\sigma-\sigma^\prime)$ the $\delta$-distribution in $p$-dimension. The $p$, for which $\sigma$, $\int$, $\delta$ and the spatial world-volume differential $\mathrm{d}$ are to be understood, should be clear from context.

The currents \eqref{eq:DefZgeneral} are characterised by a charge $\mathcal{Q}: \ \bigwedge^{p-1} \mathcal{R}_1 \rightarrow  \mathcal{R}_2$, an $\mathcal{R}_2$-valued $(p-1)$-form on the extended space, parameterised by coordinates $X^M$. We define
\begin{align*}
\bfeta_{KL} = \eta_{\mathcal{N},KL} \mathcal{Q}^{\mathcal{N}}=  \frac{1}{(p-1)!} \eta_{\mathcal{N},KL} q^{\mathcal{N}}_{N_1 ... N_{p-1}} \mathrm{d}X^{N_1} \wedge ... \wedge \mathrm{d} X^{N_{p-1}} \label{eq:DefETAgeneral}
\end{align*}
in term of constants $q^\mathcal{N}_{1 ... N_{p-1}}$ and consequentially assume that $d \mathcal{Q} = 0$ resp. $\mathrm{d} \bfeta = 0$. In such a form a charge appeared already in \cite{Sakatani:2017vbd,Sakatani:2020umt}.

Again, the aim is that the algebra of sections $\phi[X] = - \int \phi^K\left(X(\sigma)\right) \bZ_K(\sigma)$, satisfying the section condition, is the standard exceptional generalised Lie derivative
\begin{equation}
\PB{\phi_1}{\phi_2}_D = - \int \left( \phi_{1}^K \partial_K  \phi_{2}^L - \phi_{2}^K \partial_K  \phi_{1}^L - {Y^{KL}}_{MN} \phi_1^M \partial_K \phi_2^N \right) \bZ_L
\end{equation}
respectively the para-Hermitian exceptional generalised Lie derivative $\PB{\phi_1}{\phi_2}_L$ for ${P^{KL}}_{MN}$ instead of ${Y^{KL}}_{MN}$. This will be the case if we assume $\PB{\bZ_M(\sigma)}{X^N(\sigma^\prime)} = \delta_M^N \delta(\sigma - \sigma^\prime)$ and the current algebra
\begin{align}
\PB{\bZ_M(\sigma)}{\bZ_N(\sigma^\prime)}_D &= \bfeta_{MN} \wedge \mathrm{d} \delta(\sigma-\sigma^\prime) \label{eq:CurrentAlgebraDorfmanGeneral}
\end{align}
respectively
\begin{align}
\PB{\bZ_M(\sigma)}{\bZ_N(\sigma^\prime)}_L &= \mathcal{Q}^{\mathcal{L}} \wedge \left( \left( \eta_{\mathcal{L},MN} + \omega_{\mathcal{L},(MN)} \right) \frac{1}{2}\left(\mathrm{d}-\mathrm{d}^\prime\right)  + \omega_{\mathcal{L},[MN]} \frac{1}{2} \left(\mathrm{d}+\mathrm{d}^\prime\right)\right) \delta(\sigma-\sigma^\prime), \label{eq:CurrentAlgebraLieGeneral}
\end{align}
and if the charge $\mathcal{Q}^{\mathcal{M}}$ fulfils the \textit{charge condition} 
\begin{equation}
\mathcal{Q}^{\mathcal{M}} \wedge \mathrm{d}X^N \partial_N = \eta^{\mathcal{M},NP} \bZ_P \partial_N. \label{eq:ChargeCond1}
\end{equation}
In terms of the $\mathcal{Q}$ alone it is, using \eqref{eq:DefZgeneral}, given by\footnote{If multiplied by $\eta_{\mathcal{M},KL}$ the charge condition can be put in terms of the $\bfeta_{KL}$ as
\begin{equation}
\bfeta_{KL}\wedge \mathrm{d}X^M \partial_M = {Y^{MN}}_{KL} \bZ_M \partial_N = \frac{1}{p} {Y^{MN}}_{KL} \bfeta_{MP} \wedge \mathrm{X}^P \partial_N. \label{eq:ChargeCond3}
\end{equation}
As such, the condition already appeared for the string ($p=1$) in \cite{Arvanitakis:2018hfn}.}
\begin{equation}
\mathcal{Q}^{\mathcal{M}} \wedge \mathrm{d}X^N \partial_N = \frac{1}{p} \eta^{\mathcal{M},NP} \eta_{\mathcal{K},LP} \mathcal{Q}^{\mathcal{K}} \wedge \mathrm{d}X^L \partial_N. \label{eq:ChargeCond2}
\end{equation}
In order to find a solution for the charge $\mathcal{Q}^\mathcal{M}$, it seems that the section condition needs to be imposed first. Only in special cases like SL$(5)$, it seems that one can solve the charge condition first and derive a section condition from a concrete charge.

\paragraph{Connecting to the previous results for SL$(5)$.}
For the $1$- and $2$-branes we can make contact with section \ref{chap:SL5} and \cite{Arvanitakis:2018hfn}
\begin{align}
p&=1: \quad q^\mathcal{M} \equiv \mathbf{q}^\mathcal{M} \in \mathcal{R}_2 \\
p&=2: \quad q^\mathcal{M}_{M_1} = q^\mathcal{M}_{\mathcal{M}_1 \mathcal{M}_2} \equiv \delta^{\mathcal{M} \mathcal{M}^\prime}_{\mathcal{M}_1 \mathcal{M}_2} \mathbf{q}_{\mathcal{M}^\prime}, \quad \text{with} \ \mathbf{q}_{\mathcal{M}^\prime} \in \overline{\mathcal{R}}_2 = \mathcal{R}_3 \\
p&=3: \quad q^\mathcal{M}_{M_1 M_2} = \epsilon^{\mathcal{M} M_3 M_4} \epsilon_{\mathcal{N}_1 M_1 M_3} \epsilon_{\mathcal{N}_2 M_2 M_4} \mathbf{q}^{\mathcal{N}_1 \mathcal{N}_2} \quad \text{with} \ \mathbf{q}^{\mathcal{N}_1 \mathcal{N}_2} \in {\bigwedge}^2 \ \mathcal{R}_2 = \mathcal{R}_4 . \label{eq:D3charge}
\end{align}
The charge condition \eqref{eq:ChargeCond2} implies
\begin{align}
\mathbf{q}^\mathcal{M} \partial_{\mathcal{M} \mathcal{M}^\prime} = 0 \in \mathcal{R}_3 , \quad \mathbf{q}_{[\mathcal{M}_1} \partial_{\mathcal{M}_2 \mathcal{M}_3]} = 0 \in \mathcal{R}_4, \quad \mathbf{q}^{\mathcal{M}_1 \mathcal{M}_2} \partial_{\mathcal{M}_1 \mathcal{M}_2} = 0 \in \mathcal{R}_5, \label{eq:SL5ChargeConditionCondensed}
\end{align}
using the SL$(5)$ representations $\mathcal{R}_p$. Motivated by that one might suggest that also for general $d$ the charges of $p$-branes are elements of representations $\mathcal{R}_{p+1}$ appearing in the tensor hierarchy, c.f. table \ref{table:Representations}, $\mathcal{Q}_{(p)} \in \mathcal{R}_{p+1}$ and their charge conditions to be elements in $\mathcal{R}_{p+2}$ as was already done in \cite{Arvanitakis:2018hfn}. This claim is true for the SL$(5)$-theory as demonstrated above.

Apart for the $p=3$-case, D3-branes in the IIb theory, for which a similar embedding $\mathcal{Q}_{(3)} = \mathcal{R}_4 \in \bigwedge^2 \mathcal{R}_1 \otimes \mathcal{R}_2$ seems possible for arbitrary $d$, and the obvious $\mathcal{Q}_{(1)} \in \mathcal{R}_2$, there seems no necessity in our approach to the current algebra to assume that the charges or their charge condition should fit into the tensor hierarchy. We worked so far purely from the point of view of the theory in the internal space. Furthermore, the charges should exist also in cases, which are normally not considered to be part of the tensor hierarchy, as for example for the $5$-brane multiplet in the $E_{6(6)}$, for which the charges then would take values in $\mathcal{R}_6$. So, in the following, we will continue to consider the charges to be rather $\mathcal{Q}_{(p)} \in \left(\bigwedge {}^{p-1} \  \overline{\mathcal{R}}_1 \right) \otimes \mathcal{R}_2$ as before, and not elements of $\mathcal{R}_{p+1}$.

\subsection{Canonical $p$-form currents, their current algebra and SL$(p+3)$} \label{chap:pForm}
The obvious generalisation of the type II string in $d=3$ and membrane in $d=4$ cases is the $p$-brane current in a $p+2$-dimensional target space\footnote{Similar nice looking expressions can be obtained for any $d\geq p+1$, but only for $d=p+2$ the currents form a representation of an SL group, SL$(p+3)$. The occurring currents have been investigated already in \cite{Bonelli:2005ti} in the strive for topological theories.}
\begin{equation}
\bZ_M (\sigma)= \left( p_\mu(\sigma) \star 1 , \frac{1}{\sqrt{p!}} \mathrm{d} x^{\mu_1} \wedge ... \wedge \mathrm{d} x^{\mu_p} \right).
\end{equation}
Using the canonical Poisson structure, one obtains
\begin{align}
\PB{\bZ_M(\sigma)}{\bZ_N(\sigma^\prime)} &= \mathcal{Q}^{\mathcal{L}} \wedge \left( \eta_{\mathcal{L},MN} \frac{1}{2}\left(\mathrm{d}-\mathrm{d}^\prime\right)  + \omega_{\mathcal{L},MN} \frac{1}{2} \left(\mathrm{d}+\mathrm{d}^\prime\right)\right) \delta(\sigma-\sigma^\prime) \label{eq:CurrentAlgebraSL(p+3)}
\end{align}
with $\eta$- and $\omega$-symbols of the SL$(p+3)$-group \eqref{eq:SL(d)Symbols} and
\begin{equation}
\mathcal{Q}^{\mathcal{M}} = \left(\frac{1}{\sqrt{(p-1)!}} \mathcal{Q}^{\mu_1 ... \mu_{p-1}},0\right), \qquad \mathcal{Q}^{\mu_1 ... \mu_{p-1}}  = \mathrm{d}x^{\mu_1} \wedge ... \wedge \mathrm{d}x^{\mu_{p-1}}.
\end{equation}
This is a solution to the charge condition \eqref{eq:ChargeCond2} to the SL$(p+3)$ $\eta$-symbols and indeed the form of the current fits into the above system $\bZ_M = \frac{1}{p} \eta_{\mathcal{K},LM} \mathcal{Q}^K \wedge \mathrm{d}X^L$:
\begin{align}
\bZ_M = \left( \frac{1}{p!} \mathrm{d} x^{\kappa_1} \wedge .... \wedge \mathrm{d} x^{\kappa_{p-1}} \wedge \mathrm{d} \tilde{x}_{\kappa_1 ... \kappa_{p-1} \kappa} \ ,  \ \frac{1}{\sqrt{p!}} \mathrm{d} x^{\mu_1} \wedge ... \wedge \mathrm{d} x^{\mu_p} \right) \label{eq:pBraneCurrent}
\end{align}
with coordinates $\tilde{x}_{\kappa_1 ... \kappa_p}$ corresponding to the $p$-brane momentum (c.f. the string case: $p = \partial x$). When neglecting total derivative contributions \eqref{eq:CurrentAlgebraSL(p+3)} becomes
\begin{align}
\PB{\bZ_M(\sigma)}{\bZ_N(\sigma^\prime)} &=  \eta_{\mathcal{L},MN} \mathcal{Q}^{\mathcal{L}} \wedge\mathrm{d}\delta(\sigma-\sigma^\prime), \label{eq:CurrentAlgebraSL(p+3)Dorfman}
\end{align}
as wished, and gives the SL$(p+3)$ generalised Lie derivative \eqref{eq:SL(d)GenLieDerivative}, when applied on \linebreak $\phi = - \int \phi^K \bZ_K$. If, instead, one used \eqref{eq:CurrentAlgebraSL(p+3)}, one reproduces the standard Lie derivative of the para-Hermitian generalised geometry on $(T\oplus \bigwedge^p T^\star)M$ \eqref{eq:SLdLieDerivative}.

\paragraph{Dirac brackets.} As for the O$(d,d)$ string, an alternative derivation of \eqref{eq:CurrentAlgebraSL(p+3)Dorfman} exists. This again employs Dirac brackets on an extended phase space with $\mathcal{R}_1$ momenta and coordinates
\begin{align}
\PB{\mathbf{P}_M(\sigma)}{X^N(\sigma^\prime)} = - (p+1) \delta_M^N \delta(\sigma - \sigma^\prime)  \label{eq:pBraneextendedphasespacePB} \\ 
\PB{\mathbf{P}_M(\sigma)}{\mathbf{P}_N(\sigma^\prime)} = \PB{X^M(\sigma)}{X^N(\sigma^\prime)} =0 \nonumber
\end{align}
for constraints $\bPhi_M = \mathbf{P}_M - \bZ_M =  \mathbf{P}_M - \frac{1}{p} \eta_{\mathcal{K},LM} \mathcal{Q}^K \wedge \mathrm{d}X^L$. One finds, neglecting total world-volume derivatives,
\begin{align*}
A_{KL}(\sigma,\sigma^\prime) &= \PB{\bZ_K(\sigma)}{\bPhi_L(\sigma^\prime)} = \PB{\bZ_K(\sigma)}{\mathbf{P}_L(\sigma^\prime)} \\
&= (p+1) \left( \begin{array}{cc} \alpha_{\kappa\lambda} & \delta^{[\lambda_p}_\kappa \mathcal{Q}^{\lambda_1 ... \lambda_{p-1}]} \\
p \delta^{[\kappa_p}_\lambda \mathcal{Q}^{\kappa_1 ... \kappa_{p-1}]} & 0 \end{array} \right) \wedge \mathrm{d} \delta(\sigma-\sigma^\prime) \\
B_{KL}(\sigma,\sigma^\prime) &= \PB{\bPhi_K(\sigma)}{\bPhi_L(\sigma^\prime)} = - \frac{(p+1)^2}{p} \eta_{\mathcal{M},KL} \mathcal{Q}^\mathcal{M} \wedge \mathrm{d}\delta(\sigma-\sigma^\prime)
\end{align*}
with $\alpha_{\kappa\lambda} = - \alpha_{\lambda\kappa}$. The Dirac brackets on \eqref{eq:pBraneextendedphasespacePB} are, as striven for,
\begin{align}
\PB{\bZ_K(\sigma)}{\bZ_L(\sigma^\prime)}_{D.B.} &= - \int^{{\prime \prime} }\int^{{\prime \prime \prime}} A_{KM}(\sigma,\sigma^{\prime \prime} ) (B^{-1})^{MN}(\sigma^{\prime \prime},\sigma^{\prime \prime \prime})  A_{NL}(\sigma^{\prime \prime \prime},\sigma^\prime )  \nonumber \\
&= \eta_{\mathcal{M},KL} \mathcal{Q}^\mathcal{M} \wedge \mathrm{d}\delta(\sigma-\sigma^\prime) \label{eq:pBraneDiracBrackets}, \\
\PB{\bZ_M(\sigma)}{X^N(\sigma^\prime)}_{D.B} &= - \delta^N_M \delta(\sigma-\sigma^\prime). \nonumber
\end{align}

\subsection{Exceptional currents for $d\leq6$} \label{chap:ExCurrent}
The previous setting was quite restrictive and only in part related to the exceptional duality groups that we want to geometrise. In particular, one can find solutions to the charge condition \eqref{eq:ChargeCond2} for each $\frac{1}{2}$-BPS object in type IIb and eleven-dimensional supergravity and construct potential currents of world-volume theories, such that their algebras respect the exceptional symmetries.

\paragraph{Solutions to the charge condition in $E_{d(d)}$.}
In the following, solutions of the charge condition and their corresponding currents $\bZ_M = \frac{1}{p} \eta_{\mathcal{K},LM} \mathcal{Q}^{\mathcal{K}} \wedge \mathrm{d} X^L$ in an M-theory or type IIb section in $E_{d(d)}$ are collected.\footnote{The currents and charges for the \textbf{type IIa theory} objects -- F1, D2, D4, NS5 -- can be obtained dimensional reduction analogous to section \ref{chap:reductionIIa}:
\begin{itemize}
\item M2 $\rightarrow$ F1, M5 $\rightarrow$ D4: $x^d \equiv \sigma^p, \quad x^{\underline{\mu}} = x^{\underline{\mu}} (\sigma^1, ... , \sigma^{p-1}), \quad \mu = (\underline{\mu},d)$
\item M2 $\rightarrow$ D2, M5 $\rightarrow$ NS5: $\mathrm{d} x^d \equiv 0$
\end{itemize}
The point particle, $p = 0$, exists as well ($\mathcal{Q} = 0$) \cite{Sakatani:2017vbd}, but has no spatial world-volume and, hence, there is also no interesting current algebra defined on it.} In contrast to the previously discussed $p=1,2$ cases in the SL$(5)$-theory, it does not seem possible to solve the condition without specifying the section first.
\begin{itemize}
\item In the \textbf{M-theory section}, we have directly from \eqref{eq:ChargeCond1}:
\begin{align}
\bZ_M = \left( Z_\mu , \frac{1}{\sqrt{2!}} \mathcal{Q}^{[\mu_1} \wedge \mathrm{d}x^{\mu_2]}, \frac{1}{\sqrt{5!}} \mathcal{Q}^{[\mu_1 \mu_2 \mu_3 \mu_4} \wedge \mathrm{d}x^{\mu_5]} \right) \label{eq:CurrentMTheorySection}
\end{align}
\begin{itemize}
\item M2-brane:
\begin{align}
\mathcal{Q}^{\mu} &= q \ \mathrm{d}x^{\mu}, \qquad \mathcal{Q}^{\mu_1 \mu_2 \mu_3 \mu_4} = 0 \label{eq:CurrentM2}\\
\bZ_M &= q \left( \frac{1}{2!} \mathrm{d} x^{\lambda} \wedge \mathrm{d} \tilde{x}_{\lambda \mu} \ , \ \frac{1}{\sqrt{2!}} \mathrm{d}x^{\mu_1} \wedge \mathrm{d}x^{\mu_2} \ , \  0 \  \right)  \nonumber
\end{align}
\item M5-brane:
\begin{align}
\mathcal{Q}^{\mu} &= -\frac{q}{6} \ \mathrm{d}x^{\mu} \wedge \mathrm{d}x^{\mu_1} \wedge \mathrm{d}x^{\mu_2} \wedge \mathrm{d} \tilde{x}_{\mu_1 \mu_2} \nonumber \\
\mathcal{Q}^{\mu_1 \mu_2 \mu_3 \mu_4} &=  \ q \ \mathrm{d}x^{\mu_1} \wedge \mathrm{d}x^{\mu_2} \wedge \mathrm{d}x^{\mu_3} \wedge \mathrm{d}  x^{\mu_4} \label{eq:CurrentM5}\\
\bZ_M &= q \left( \frac{1}{5!} \mathrm{d} x^{\mu_1} \wedge ... \wedge \mathrm{d} x^{\mu_4} \wedge \mathrm{d} \tilde{x}_{\mu_1 ... \mu_4 \mu} - \frac{1}{6 \cdot 5} \mathrm{d} x^{\mu_1} \wedge ... \wedge \mathrm{d} x^{\mu_3} \wedge \mathrm{d} \tilde{x}_{\mu_1 \mu_2} \wedge \mathrm{d} \tilde{x}_{\mu_3 \mu} \right. \nonumber \\
&{} \quad \left. \frac{1}{\sqrt{2!}} \frac{1}{6} \mathrm{d}x^{\mu_1} \wedge \mathrm{d}x^{\mu_2} \wedge \mathrm{d}x^{\mu_3} \wedge \mathrm{d}x^{\mu_4} \wedge \mathrm{d} \tilde{x}_{\mu_3 \mu_4}  \ , \ \frac{1}{\sqrt{5!}} \mathrm{d}x^{\mu_1} \wedge ... \wedge \mathrm{d}  x^{\mu_5} \right)  \nonumber
\end{align}
\end{itemize}

\item In the \textbf{type IIb section} one derives from \eqref{eq:ChargeCond1}:
\begin{align}
\bZ_M = \left( Z_{\underline{\mu}} , \mathcal{Q}_m \wedge \mathrm{d} x^{\underline{\mu}}, \frac{1}{\sqrt{3!}} \mathcal{Q}^{[\underline{\mu}_1 \underline{\mu}_2} \wedge \mathrm{d} x^{\underline{\mu}_3]} , \frac{1}{\sqrt{5!}} \mathcal{Q}^{[{\underline{\mu}_1} {\underline{\mu}_2} {\underline{\mu}_3} {\underline{\mu}_4}}_m \wedge \mathrm{d}x^{{\underline{\mu}_5}]} \right) \label{eq:CurrentIIBSection}
\end{align}
\begin{itemize}
\item F1/D1-system resp. $(p,q)$-string
\begin{align}
\mathcal{Q}_m &= q_m = \left( \begin{array}{c} p \\ q \end{array} \right), \qquad \mathcal{Q}^{\underline{\mu}_1 \underline{\mu}_2} = 0, \qquad \mathcal{Q}^{{\underline{\mu}_1} {\underline{\mu}_2} {\underline{\mu}_3} {\underline{\mu}_4}}_m = 0 \label{eq:CurrentF1D1} \\
\bZ_M &= q_m \left( \mathrm{d} \tilde{x}^m_{\underline{\mu}} \ , \ \mathrm{d}x^{\underline{\mu}} \ , \  0 \ , \ 0 \right)  \nonumber
\end{align}
\item D3-brane:\footnote{As suggested by \eqref{eq:D3charge} there might be an embedding of $\mathcal{R}_4 = \bigwedge^2 \mathcal{R}_2$ of this charge solution. This is indeed the case - the only non-zero component of $\mathbf{q}^{\mathcal{M}_1 \mathcal{M}_2}$ is $\mathbf{q}_{m_1 m_2} = \frac{1}{2} \ q \ \epsilon_{m_1 m_2}$ in the IIb section. With \eqref{eq:D3charge} one reproduces $\mathcal{Q}^\mathcal{M}$ of the D3-brane.}
\begin{align}
\mathcal{Q}_m &= \frac{1}{2} \ q \ \epsilon_{mn} \mathrm{d}x^{\underline{\mu}} \wedge \mathrm{d} \tilde{x}^n_{\underline{\mu}}, \quad \mathcal{Q}^{\underline{\mu}_1 \underline{\mu}_2} =  q \ \mathrm{d} x^{\underline{\mu}} \wedge \mathrm{d} x^{\underline{\nu}}, \quad \mathcal{Q}^{{\underline{\mu}_1} {\underline{\mu}_2} {\underline{\mu}_3} {\underline{\mu}_4}}_m = 0 \nonumber \\
\bZ_M &= q \left( \frac{1}{3!} \epsilon_{mn} \mathrm{d} x^{\underline{\nu}} \wedge \mathrm{d} \tilde{x}_{\underline{\nu}}^m \wedge \mathrm{d} \tilde{x}_{\underline{\mu}}^n - \frac{1}{3!} \mathrm{d} x^{\underline{\nu}_1} \wedge \mathrm{d} x^{\underline{\nu}_2} \wedge \mathrm{d} \tilde{x}_{\underline{\mu}_1 \underline{\mu}_2 \underline{\mu}}, \right. \label{eq:CurrentD3} \\
&{} \qquad \left. \frac{1}{2} \epsilon_{mn} \mathrm{d} x^{\underline{\mu}} \wedge \mathrm{d} x^{\underline{\nu}} \wedge \mathrm{d} \tilde{x}_{\underline{\mu}}^n  \ , \ \frac{1}{\sqrt{3!}} \mathrm{d}x^{\underline{\mu}_1} \wedge \mathrm{d}x^{\underline{\mu}_2} \wedge \mathrm{d}x^{\underline{\mu}_3} \ , \ 0 \right)  \nonumber
\end{align}
\item NS5/D5-brane:
\begin{align}
\mathcal{Q}_m &= \frac{1}{4!} q_l \left( 3 \ \epsilon_{mn} \mathrm{d}x^{\underline{\nu}} \wedge \mathrm{d} \tilde{x}^n_{\underline{\nu}} \wedge \mathrm{d}x^{\underline{\lambda}} \wedge \mathrm{d} \tilde{x}^l_{\underline{\lambda}} - \delta^l_m \mathrm{d} x^{\underline{\mu}_1} \wedge \mathrm{d} x^{\underline{\mu}_2} \wedge  \mathrm{d} x^{\underline{\mu}_3} \wedge \mathrm{d} \tilde{x}_{\underline{\mu}_1 \underline{\mu}_2 \underline{\mu}_3} \right) \nonumber\\
 \mathcal{Q}^{\underline{\mu}_1 \underline{\mu}_2} &=  \frac{1}{2} \ q_m \ \mathrm{d} x^{\underline{\mu}_1} \wedge \mathrm{d} x^{\underline{\mu}_2} \wedge  \mathrm{d} x^{\underline{\mu}_3} \wedge \mathrm{d} \tilde{x}_{\underline{\mu}_3}^m \nonumber \\
 \mathcal{Q}^{{\underline{\mu}_1} {\underline{\mu}_2} {\underline{\mu}_3} {\underline{\mu}_4}}_m &= \ q_m \ \mathrm{d} x^{\underline{\mu}_1} \wedge \mathrm{d} x^{\underline{\mu}_2} \wedge  \mathrm{d} x^{\underline{\mu}_3} \wedge \mathrm{d} x^{\underline{\mu}_4}  \label{eq:CurrentNS5D5} \\
\bZ_M &= \left(  \frac{1}{5} \left[\mathcal{Q}_m \wedge \mathrm{d} \tilde{x}^m_{\underline{\mu}} + \frac{1}{2!} \mathcal{Q}^{\underline{\mu}_1 \underline{\mu}_2} \wedge \mathrm{d} \tilde{x}_{\underline{\mu}_1 \underline{\mu}_2 \underline{\mu}} + \frac{1}{4!} \mathcal{Q}^{\underline{\mu}_1 \underline{\mu}_2 \underline{\mu}_3 \underline{\mu}_4}_m \wedge \mathrm{d} \tilde{x}^m_{\underline{\mu}_1 \underline{\mu}_2 \underline{\mu}_3 \underline{\mu}_4 \underline{\mu}_5}  \right] \ ,  \right. \nonumber \\
&{} \quad  - \frac{1}{4!} \ q_l \ \mathrm{d}x^{\underline{\mu}} \wedge \mathrm{d}x^{\underline{\mu}_1} \wedge \mathrm{d}x^{\underline{\mu}_2} \wedge \left( 3 \ \epsilon_{mn} \mathrm{d} \tilde{x}^n_{\underline{\mu}_1} \wedge \mathrm{d} \tilde{x}^l_{\underline{\mu}_2} + \delta^l_m  \mathrm{d} x^{\underline{\mu}_3} \wedge \mathrm{d} \tilde{x}_{\underline{\mu}_1 \underline{\mu}_2 \underline{\mu}_3} \right) , \nonumber \\
&{} \quad \left. \frac{1}{\sqrt{3!}} \frac{1}{2} \ q_m \ \mathrm{d} x^{\underline{\mu}_1} \wedge ... \wedge \mathrm{d} x^{\underline{\mu}_4} \wedge \mathrm{d} \tilde{x}_{\underline{\mu}_4}^m \ , \ \frac{1}{\sqrt{5!}} \ q_m \ \mathrm{d} x^{\underline{\mu}_1} \wedge ... \wedge \mathrm{d} x^{\underline{\mu}_5}  \right)  \nonumber
\end{align}
\end{itemize}
\end{itemize}
Suggested by these charge solutions, we see the typical decomposition of $\mathcal{R}_1$ and $\mathcal{R}_2$ in the M-theory and type IIb sections
\begin{align}
\bZ_M &= \Big(Z_\mu,\underbrace{\frac{\bZ^{\mu_1 \mu_2}}{\sqrt{2!}}}_{\text{M2}},\underbrace{\frac{\bZ^{\mu_1 \mu_2 \mu_3 \mu_4 \mu_5}}{\sqrt{5!}}}_{\text{M5}} \Big), \qquad \mathcal{Q}^\mathcal{M} = \Big( \underbrace{Q^\mu}_{\text{M2}} , \underbrace{\frac{Q^{\mu_1 \mu_2 \mu_3 \mu_4}}{\sqrt{4!}}}_{\text{M5}},\underbrace{\frac{Q^{\mu_1 \mu_2 \mu_3 \mu_4 \mu_5 \mu_6,\mu}}{\sqrt{6!}}}_{\text{KKM}} \Big) \\
\bZ_M &= \Big(Z_\mu,\underbrace{\bZ_m^{\underline{\mu}}}_{\text{F1/D1}},\underbrace{\frac{\bZ^{{\underline{\mu}}_1 {\underline{\mu}}_2 {\underline{\mu}}_3}}{\sqrt{3!}}}_{\text{D3}},\underbrace{\frac{\bZ_m^{{{\underline{\mu}}_1 {\underline{\mu}}_2 {\underline{\mu}}_3 {\underline{\mu}}_4 {\underline{\mu}}_5}}}{\sqrt{5!}}}_{\text{NS5/D5}} \Big), \quad \mathcal{Q}^\mathcal{M} = \Big(  \underbrace{Q_m}_{\text{F1/D1}}, \underbrace{\frac{Q^{\underline{\mu}_1 {\underline{\mu}}_2}}{\sqrt{2!}}}_{\text{D3}} , \underbrace{\frac{Q^{{\underline{\mu}}_1 {\underline{\mu}}_2 {\underline{\mu}}_3 {\underline{\mu}}_4}_m}{\sqrt{4!}}}_{\text{NS5/D5}} , \underbrace{\frac{Q^{{\underline{\mu}}_1 {\underline{\mu}}_2 {\underline{\mu}}_3 {\underline{\mu}}_4 {\underline{\mu}}_5,{\underline{\mu}}}}{\sqrt{5!}}}_{\text{KKM}}  \Big) \nonumber
\end{align}
The already existing place for the Kaluza-Klein monopole (KKM) in the $\mathcal{R}_2$-representation is empty in the above collected list of currents. It is simply not a solution to the charge condition \eqref{eq:ChargeCond2} for $d\leq6$ yet. The decomposition of the $\bZ$ is similar to the one of the generalised coordinates in $X^M$. For both sections the $x^\mu$ and $Z_\mu = p_\mu$-components are shared in all currents.

The \textit{non-geometric coordinates} of $X^M$, i.e. the ones that do not belong to a section, could be associated to the momenta of the corresponding objects -- in the same decomposition as the one for the $\bZ$. This is analogous to the string in O$(d,d)$ generalised geometry, where $p_\mu = \mathrm{d} \tilde{x}_\mu$ \cite{Blair:2013noa,Blair:2014kla,Osten:2019ayq}.

\paragraph{The non-geometry of the currents.} As already noted previously \cite{Sakatani:2017vbd,Sakatani:2020umt}, where $\sigma$-model constructions of such world-volume theories were considered, we observe that for the 'higher' ($p>2$) world-volume not only the corresponding $Q$-component but also all the 'lower' ones are turned on -- e.g. for the M5-brane \eqref{eq:CurrentM5} not only $\mathcal{Q}^{\mu_1 ... \mu_4}$ but also $\mathcal{Q}^{\mu}$. In the end, the reason for that is $E_{d(d)}$-covariance.

In \cite{Sakatani:2017vbd,Sakatani:2020umt} and, from a current algebra approach for the M5-brane, \cite{Hatsuda:2013dya} this was done by introducing the gauge fields, that typical for M5-brane actions \cite{Pasti:1997gx,Aganagic:1997zq}. Here, instead, these components contain differentials of \textit{non-geometric coordinates}, that are associated to the momenta of the lower dimensional branes. In the example of the NS/D5-brane \eqref{eq:CurrentNS5D5}, the D3-charge and current contain the differential $\mathrm{d}\tilde{x}^m_{\underline{\mu}_4}$ and so on. Hence, one interpretation of the extended coordinates could be to account for these d.o.f.s.

Consequentially, from the point of view of such a higher brane, whose phase space coordinate normally just would be $x^\mu$ and $p_\mu = Z_\mu$, the currents itself already are non-geometric as also the additional non-geometric coordinates $\tilde{x}$ appear inevitably. Finally, such terms are logical to appear if one strives for an $E_{d(d)}$-covariant setup: As for example the M2- and M5-brane or the D-branes are connected by duality, in general M5 currents can't appear alone in such a setup. Another way to understand that fact is, that in the generalised Lie derivative for $d>4$ (in the M-theory picture) a 'coupling' of two-forms appears: $[\omega_1 , \omega_2 ]_D = \omega_1 \wedge \mathrm{d} \omega_2$ for two-forms $\omega_i$. Hence, the current algebra bracket of two M2-currents forms an M5-current, necessitating that in presence of an M5-current also M2-currents have to be present. Interestingly, the converse does not seem to be true: one can formulate an M2-brane current algebra, by consistently setting the (canonical) M5-current to zero.

\paragraph{Deriving the current algebra?} As the generalised Lie derivative for $E_{d(d)}$ for $d\geq 5$ is not equivalent to the standard Lie derivative up to total derivatives, \textit{one cannot hope to derive this current algebra} in a similar way to the $p$-brane in SL$(p+3)$ or the string in O$(d,d)$. Neither starting from canonical Poisson brackets nor the $E_{d(d)}$-invariant Dirac bracket approach lead to a non-Lie bracket. The only reason, why one could make it work in the previous cases, is that there the difference is always only a total-derivative term on the spatial world-volume, that would not contribute classically.

It is possible to derive a para-Hermitian $E_{d(d)}$ current, though. For example, for $d=5,6$, it could simply be realised as a 'sum' of the $p=2$- and $p=5$-currents resp. their corresponding SL$(p+3)$-theory
\begin{equation}
\bZ_M = \bZ^{(2)}_M(\sigma) + \bZ^{(5)}_M(\rho) \sim \left( p^{(2)}_\mu(\sigma) + p^{(5)}_\mu(\rho), \mathrm{d} x^{\mu_1}(\sigma) \wedge \mathrm{d} x^{\mu_2}(\sigma), \mathrm{d} x^{\mu_1}(\rho) \wedge ... \wedge \mathrm{d} x^{\mu_5}(\rho) \right) \nonumber
\end{equation}
in the M-theory section. They correspond to decoupled M2 and M5 world-volume theories, without exhibiting the exceptional symmetry at the classical level.

\paragraph{Comparison to the Hatsuda-Kamimura M5-current.} As a consistency check, we compare the M5-currents \eqref{eq:CurrentM5} to the M5-current of the SO$(5,5)$ theory in \cite{Hatsuda:2013dya} that were shown to encode the Hamiltonian formulation of the Pasti-Sorokin-Tonin M5-brane action \cite{Pasti:1997gx,Aganagic:1997zq}.

If we make the identifications
\begin{equation}
p_\mu = \bZ_\mu \quad \text{and} \quad F = \frac{1}{6} \mathrm{d} x^\mu \wedge \mathrm{d} x^{\mu^\prime} \wedge \mathrm{d} \tilde{x}_{\mu \mu^\prime}
\end{equation}
in \eqref{eq:CurrentM5}, we reproduce the currents from \cite{Hatsuda:2013dya}. The charge corresponding to this current is the one from \cite{Sakatani:2017vbd,Sakatani:2020umt}
\begin{equation}
\mathcal{Q}_{[F]}^\mathcal{M} = \left( F \wedge \mathrm{d}x^\mu \ , \ \mathrm{d}x^{\mu_1} \wedge ... \wedge \mathrm{d} x^{\mu_4} , 0 \right) \label{eq:IdentificationsHatsudaKamimura}
\end{equation}
when plugged into \eqref{eq:DefZgeneral}. From \eqref{eq:CurrentAlgebraDorfmanGeneral} we derive the formal expression
\begin{equation}
\PB{F(\sigma)}{F(\sigma^\prime)} = \mathrm{d}\delta(\sigma - \sigma^\prime)
\end{equation}
or $\PB{\bar{F}^{\alpha \alpha^\prime}(\sigma)}{\bar{F}^{\beta \beta^\prime}(\sigma^\prime)} = \epsilon^{\alpha \alpha^\prime \beta \beta^\prime \gamma} \partial_\gamma \delta(\sigma - \sigma^\prime)$ for $F_{\alpha_1 \alpha_2 \alpha_3} = \frac{1}{2} \epsilon_{\alpha_1 \alpha_2 \alpha_3 \beta_1 \beta_2} \bar{F}^{\beta_1 \beta_2} \delta(\sigma - \sigma^\prime)$. As shown in \cite{Hatsuda:2013dya}, this is indeed the Dirac bracket of a self-dual two-form gauge field $A$ with field strength $F = \mathrm{d}A$ (on the M5-brane world-volume) and canonical momentum $E$ -- $\PB{E^{\alpha \alpha^\prime}(\sigma)}{A_{\beta \beta^\prime}(\sigma^\prime)} = - \delta^{\alpha}_{[\beta} \delta^{\alpha^\prime}_{\beta^\prime]}$ -- w.r.t. to the constraints $\Phi = E-\bar{F} = 0$. Hence, one can derive from the canonical phase space $\left(p_\mu(\sigma) , x^\mu(\sigma); E(\sigma),A(\sigma) \right)$ a current algebra
\begin{equation}
\PB{\bZ_{M}(\sigma)}{\bZ_N(\sigma^\prime)}_{[F]} = \mathcal{Q}_{[F]}^\mathcal{L} \wedge \left( \eta_{\mathcal{L},MN} \frac{1}{2}\left(\mathrm{d}-\mathrm{d}^\prime\right)  + \omega_{\mathcal{L},[MN]} \frac{1}{2} \left(\mathrm{d}+\mathrm{d}^\prime\right)\right) \delta(\sigma-\sigma^\prime).
\end{equation}
without the previously unwanted $\omega_{\mathcal{L},(MN)}$-terms and, hence, it corresponds to the striven for Dorfman current algebra up to total world-volume derivative terms. However, it does only satisfy the Jacobi identity up to such total-derivative terms and, resultantly, does not correspond to para-Hermitian exceptional geometry.

Of course, another price for deriving the Dorfman current algebra \eqref{eq:CurrentAlgebraDorfmanGeneral} here was the loss of $E_{d(d)}$-invariance, due to the identifications \eqref{eq:IdentificationsHatsudaKamimura}. The formalism presented in this paper allows encode such degrees as the gauge field $A$ in an $E_{d(d)}$-covariant form in form of the extended coordinates $\tilde{x}_{\mu \mu^\prime}$. One could expect that the gauge field momentum $E$ is a (non-local) incarnation of the extended momentum $P^{\mu \mu^\prime}$.

\section{Outlook}
There are several conceptual questions left open in this article. The main result, the $E_{d(d)}$-covariant current algebra
\begin{equation}
\PB{\bZ_M(\sigma)}{\bZ_N(\sigma^\prime)} = \eta_{\mathcal{L},MN} \mathcal{Q}^\mathcal{L} \wedge \mathrm{d}\delta(\sigma - \sigma^\prime) \label{eq:OutlookCurrent}
\end{equation}
of a $p$-brane current was postulated but could not be derived from some underlying principle, like a canonical (Poisson) structure, in general, when also imposing a $E_{d(d)}$-invariant definition of the current $\bZ$. It would be nice to change that by considering higher algebraic structures, like aspects of dg-symplectic geometry, $L_{\infty}$ algebras or AKSZ constructions, see for example \cite{Arvanitakis:2018cyo} for similar approaches in the context of exceptional generalised geometry. On a more fundamental level, one might ask whether a canonical phase space structure could ever capture the symmetries of M-theory, or whether on should take a structure like \eqref{eq:OutlookCurrent} to be the fundamental one.

The current algebra was constructed as a simple way to write down an $E_{d(d)}$-invariant world-volume theory, motivated by the membrane and string in the SL$(5)$-theory where it could be derived from the standard Polyakov type action, when neglecting total world-sheet derivatives. A similar clean path to an action could be possible for D-branes, following \cite{Hatsuda:2012uk}. For $p>2$ the difference between the \textit{canonical} and the $E_{d(d)}$-current algebra is not a total world-volume differential anymore (as in the string and the membrane case). This relates directly to discussions at the level of the action, where such topological contributions have been shown to appear as well, as difference between duality invariant and standard versions of actions \cite{Tseytlin:1990nb,Hull:2004in,Hull:2006qs,Berman:2007vi}. Hence, for $p>2$ severe modification of actions in contrast to the normal Polyakov Nambu-Goto type ones are to be expected, when demanding $E_{d(d)}$ covariance. Several candidates of such $E_{d(d)}$-invariant world-volume actions have been discussed in the literature \cite{Sakatani:2016sko,Blair:2017hhy,Sakatani:2017vbd,Arvanitakis:2018hfn,Sakatani:2020umt}. It would be interesting to study whether they would correspond to the Hamiltonian settings presented in this paper. This would also clarify, whether the conjectured Hamiltonian and spatial diffeomorphism
\begin{align}
H &= \frac{1}{2} \int \bZ_K \wedge \star \bZ_L \mathcal{H}^{KL}, \qquad 0 = \eta^{\mathcal{M}, K L} \bZ_K \wedge \star \bZ_L. \label{eq:OutlookHamiltonian}
\end{align}
make sense for $p>2$ and $d>4$ as well on the extended space. When including gauge fields instead of dual coordinates this has been demonstrated for the M5-brane for $d=5$ \cite{Hatsuda:2013dya}.n The Hamiltonian formulation of membranes in a duality invariant way might be helpful in context of the recent study of exceptional Drinfeld algebras \cite{Sakatani:2019zrs,Malek:2019xrf,Sakatani:2020iad,Bakhmatov:2020kul,Blair:2020ndg,Musaev:2020bwm,Malek:2020hpo,Sakatani:2020wah,Gubarev:2020ydf,Musaev:2020nrt}, in particular in the question whether certain transformations, e.g. polyvector shifts a.k.a. generalised Yang-Baxter deformations, of Drinfeld algebras really correspond to dualities of some world-volume theory, generalising work of \cite{Sakatani:2020iad} to generic objects and $d>4$. But one should not forget that the current algebra \eqref{eq:OutlookCurrent} is, in general, not a Lie algebra. Tools of conventional classical field theory, like Legendre transformations or canonical transformations, have to be reconsidered. Generalising techniques of geometric quantisation from the recently studies O$(d,d)$ case \cite{Alfonsi:2021bot} could be necessary, as well.

The fact, that all the $p>2$ exceptional currents in section \ref{chap:ExCurrent} are manifestly non-geometric, is a puzzling results. Perhaps one must consider multiple interacting world-volume theories at once, e.g. M2- and M5-branes, to make sense out of that Hamiltonian theory. Potentially this might help resolving the famous membrane duality problem \cite{Duff:1989tf,Duff:2015jka}. In the string case, the study of the current algebra was feasible in order interpret backgrounds as non-commutative and non-associative ones \cite{Blair:2014kla,Osten:2019ayq}. Integrating \eqref{eq:OutlookCurrent} could give a similar picture for M-theory backgrounds. The picture should be more involved, as one would expect all kinds of higher brackets to appear. 

Generalisations of the present setup to $E_{7(7)}$ and $E_{8(8)}$ generalised geometry should be possible as well. The additional duality group invariants, that appear in the generalised Lie derivatives there, have to be introduced into the current algebra \eqref{eq:OutlookCurrent}. Extending the analysis to include the exterior space and tensor hierarchies seems straightforward.

For the membrane in the SL$(5)$-theory it was recently shown\cite{Strickland-Constable:2021afa} that the solution to the classical world-volume theory can be interpreted as generalised geodesics in SL$(5)$ generalised geometry. Studying the equations of motion for the $E_{d(d)}$- and SL$(p+3)$-currents with the above Hamiltonian, might extend that picture to higher dimensional objects and higher space-time dimension. This might give a geometric interpretation of the charge $\mathcal{Q}$. So far, it only appears as a auxiliary object to ensure manifest $E_{d(d)}$-covariance.

\section*{Acknowledgements}
I thank Alex Arvanitakis, Chris Blair and Daniel Thompson for discussions at the start of this project, and Alex Arvanitakis, Eoin O Colgain and Michael Duff for comments on the first version. Also, I thank the Max-Planck-Institute for Physics Munich where a large portion of this work was performed. 

\appendix
\section{Conventions and identities} \label{chap:Conventions}
We define the generalised Kronecker symbols as follows:
\begin{equation}
\epsilon^{i_1 ... i_k j_1 ... j_{n-k}} \epsilon_{i_1 ... i_k k_1 ... k_{n-k}} = k ! \ \delta^{j_1 ... j_{n-k}}_{k_1 ... k_{n-k}} = k! \ (n-k)! \ \delta^{j_1}_{[k_1} ... \delta^{j_{n-k}}_{k_{n-k}]} \nonumber
\end{equation}

The standard $\delta$-distribution $f(\sigma) = \int \mathrm{d} \sigma^\prime f(\sigma^\prime) \delta(\sigma - \sigma^\prime)$
behaves in a maybe unexpected way, when applied to functions on compact spaces. In particular $(\partial_1 + \partial_2) \delta(\sigma_2 - \sigma_1) \neq 0$,
but instead
\begin{equation}
\int \mathrm{d} \sigma_1 \mathrm{d} \sigma_2 \ \phi(\sigma_1,\sigma_2) (\partial_1 + \partial_2) \delta(\sigma_2 - \sigma_1) = \int \mathrm{d} \sigma \ \partial \left( \phi(\sigma,\sigma) \right).
\end{equation}
As discussed in the main text, this term is not vanishing in general. Strings and membranes can have non-trivial winding around a non-trivial cycle in target space. In that case, the coordinate fields $x(\sigma)$ are not smooth, such that in particular $\int \mathrm{d} \sigma \ \partial x(\sigma) \neq 0$. For open world-volumes there are boundary contributions, as well.

In many calculations it is helpful to write
\begin{equation}
\delta(\sigma_2 - \sigma_1) = \int \mathrm{d} \sigma \delta(\sigma - \sigma_1) \delta(\sigma - \sigma_2),
\end{equation}
to see for example that $(\partial_1 + \partial_2) \delta(\sigma_2 - \sigma_1) = \int \mathrm{d} \sigma \partial \delta (\sigma-\sigma_1) \delta(\sigma - \sigma_2).$ The following distributional identities follow:
\begin{align}
& \int \mathrm{d} \sigma f(\sigma) (\partial_1 + ... + \partial_n) \big(\delta(\sigma - \sigma_1) \cdot ... \cdot \delta(\sigma - \sigma_n) \big) \nonumber \\
&{} \qquad = \int \mathrm{d} \sigma \ \big( (\partial f(\sigma)) + (n-1) f(\sigma) \partial\big) \big( \delta(\sigma - \sigma_1) \cdot ... \cdot \delta(\sigma - \sigma_n) \big) \nonumber \\
& \frac{1}{2}e(\sigma_1)\cdot e^{-1}(\sigma_2) (\partial_1 - \partial_2) \big(\delta(\sigma - \sigma_1) \delta(\sigma - \sigma_2) \big) \label{eq:DistributionalIdentities}  \\
&{} \qquad = \frac{1}{2} (\partial_1 - \partial_2) \big(\delta(\sigma - \sigma_1) \delta(\sigma - \sigma_2) \big) \mathbbm{1} - \big((\partial e) \cdot e^{-1}\big)(\sigma) \delta(\sigma - \sigma_1)\delta(\sigma - \sigma_2)   \nonumber \\
& \big( f(\sigma_2) \partial_1 + f(\sigma_1) \partial_2 \big) \delta (\sigma - \sigma_1)\delta(\sigma - \sigma_2) \nonumber \\
 &{} \qquad = \big(\partial f(\sigma) \big) \delta(\sigma - \sigma_1) \delta(\sigma - \sigma_2) + \partial\big(f(\sigma) \delta(\sigma - \sigma_1) \delta(\sigma - \sigma_2)\big) \nonumber 
\end{align}
for arbitrary (matrix-valued) functions $e$ and $f$ which hold without any additional boundary terms.

The higher world-volume currents are written directly as forms (on the spatial world-volume). Typical important identities are
\begin{align*}
\mathrm{d} x^\mu (\sigma) \wedge \mathrm{d} \delta(\sigma-\sigma^\prime) &= \mathrm{d} x^\mu (\sigma^\prime) \wedge \mathrm{d} \delta(\sigma-\sigma^\prime) \\
\omega (\sigma^\prime) \wedge \mathrm{d} \delta(\sigma-\sigma^\prime) &= \omega (\sigma) \wedge \mathrm{d} \delta(\sigma-\sigma^\prime) + (\mathrm{d}\omega)(\sigma) \delta(\sigma - \sigma^\prime) \quad \text{for $(p-1)$-forms }\omega.
\end{align*}

\section{The $E_{d(d)}$ $\eta$- and $\omega$-symbols for $d\leq 6$} \label{chap:AppendixEdd}
For $d \leq 6$ the $Y$-tensor can be written in terms of the $\eta$-symbols:
\begin{equation}
{Y^{KL}}_{MN} = \eta^{\mathcal{P},KL} \eta_{\mathcal{P},MN}
\end{equation}
when we choose the normalisation
\begin{equation}
\eta^{\mathcal{M},KL} \eta_{\mathcal{N},KL} = 2(d-1)\delta^{\mathcal{M}}_{\mathcal{N}}.
\end{equation}
Indices $K,L,M,...$ resp. $\mathcal{K},\mathcal{L},\mathcal{M},...$ denote $\mathcal{R}_1$- resp. $\mathcal{R}_2$-indices of the duality group $E_{d(d)}$.

In the following, we give the $\eta$- and $\omega$-symbols explicitly in the conventions in which contractions of the totally antisymmetric multi-indices ${}^{\mu_1 ... \mu_p}$ receiving a factor $\frac{1}{p!}$. Let us note, that for $d>4$ the $\omega$-symbols are not skewsymmetric anymore: $\omega_{\mathcal{M},KL} \neq - \omega_{\mathcal{M},LK}$.

\begin{itemize}
\item M-theory decomposition with target space indices $\kappa,\lambda,\mu,... = 1,...,d$:
\begin{align*}
X^M &= \left(x^\mu,\frac{\tilde{x}_{\mu_1 \mu_2}}{\sqrt{2!}},\frac{\tilde{x}_{\mu_1 \mu_2 \mu_3 \mu_4 \mu_5}}{\sqrt{5!}} \right) \quad \in \mathcal{R}_1 \\
\mathcal{Q}^\mathcal{M} &= \left( Q^\mu , \frac{Q^{\mu_1 \mu_2 \mu_3 \mu_4}}{\sqrt{4!}},\frac{Q^{\mu_1 \mu_2 \mu_3 \mu_4 \mu_5 \mu_6,\mu}}{\sqrt{6!}} \right) \quad \in \mathcal{R}_2
\end{align*}

\begin{align*}
\eta_{\mathcal{M},KL}: \quad \qquad \eta_{\mu,KL} &= \left( \begin{array}{ccc} 0 & \frac{\delta_{\mu \kappa}^{\lambda_1 \lambda_2}}{\sqrt{2!}} & 0 \\ \frac{\delta_{\mu \lambda}^{\kappa_1 \kappa_2}}{\sqrt{2!}} & 0 & 0 \\ 0 & 0 & 0 \end{array} \right) \\
\quad \eta_{\mu_1 \mu_2 \mu_3 \mu_4,KL} &= \left( \begin{array}{ccc} 0 & 0 & \frac{\delta_{\mu_1 \mu_2 \mu_3 \mu_4 \kappa}^{\lambda_1 \lambda_2 \lambda_3 \lambda_4 \lambda_5}}{\sqrt{5!}} \\ 0 & \frac{\delta^{\kappa_1 \kappa_2 \lambda_1 \lambda_2}_{\mu_1 \mu_2 \mu_3 \mu_4}}{\sqrt{2!} \sqrt{2!}} & 0 \\ \frac{\delta_{\mu_1 \mu_2 \mu_3 \mu_4 \lambda}^{\kappa_1 \kappa_2 \kappa_3 \kappa_4 \kappa_5}}{\sqrt{5!}} & 0 & 0 \end{array} \right) \\
\left( \eta_{\mu_1 \mu_2 \mu_3 \mu_4 \mu_5 \mu_6,\mu}\right)_{KL} &= \left( \begin{array}{ccc} 0 & 0 & 0 \\ 0 & 0 & \frac{\delta_{\mu_1 \mu_2 \mu_3 \mu_4 \mu_5 \mu_6}^{\lambda_1 \lambda_2 \lambda_3 \lambda_4 \lambda_5 \nu} \delta^{\nu \mu}_{\kappa_1 \kappa_2} }{\sqrt{5!}{\sqrt{2!}}} \\
0 & \frac{\delta_{\mu_1 \mu_2 \mu_3 \mu_4 \mu_5 \mu_6}^{\kappa_1 \kappa_2 \kappa_3 \kappa_4 \kappa_5 \nu} \delta^{\nu \mu}_{\lambda_1 \lambda_2} }{\sqrt{5!}{\sqrt{2!}}} & 0
\end{array} \right) \\
\omega_{\mathcal{M},KL}: \quad \qquad \omega_{\mu,KL} &= \left( \begin{array}{ccc} 0 & - \frac{\delta_{\mu \kappa}^{\lambda_1 \lambda_2}}{\sqrt{2!}} & 0 \\ \frac{\delta_{\mu \lambda}^{\kappa_1 \kappa_2}}{\sqrt{2!}} & 0 & 0 \\ 0 & 0 & 0 \end{array} \right) \\
\quad \omega_{\mu_1 \mu_2 \mu_3 \mu_4,KL} &= \left( \begin{array}{ccc} 0 & 0 & - \frac{\delta_{\mu_1 \mu_2 \mu_3 \mu_4 \kappa}^{\lambda_1 \lambda_2 \lambda_3 \lambda_4 \lambda_5}}{\sqrt{5!}} \\ 0 & - \frac{\delta^{\kappa_1 \kappa_2 \lambda_1 \lambda_2}_{\mu_1 \mu_2 \mu_3 \mu_4}}{\sqrt{2!} \sqrt{2!}} & 0 \\ \frac{\delta_{\mu_1 \mu_2 \mu_3 \mu_4 \lambda}^{\kappa_1 \kappa_2 \kappa_3 \kappa_4 \kappa_5}}{\sqrt{5!}} & 0 & 0 \end{array} \right) \\
\left( \omega_{\mu_1 \mu_2 \mu_3 \mu_4 \mu_5 \mu_6,\mu}\right)_{KL} &= \left( \begin{array}{ccc} 0 & 0 & 0 \\ 0 & 0 & - \frac{\delta_{\mu_1 \mu_2 \mu_3 \mu_4 \mu_5 \mu_6}^{\lambda_1 \lambda_2 \lambda_3 \lambda_4 \lambda_5 \nu} \delta^{\nu \mu}_{\kappa_1 \kappa_2} }{\sqrt{5!}{\sqrt{2!}}}  \\
0 & - \frac{\delta_{\mu_1 \mu_2 \mu_3 \mu_4 \mu_5 \mu_6}^{\kappa_1 \kappa_2 \kappa_3 \kappa_4 \kappa_5 \nu} \delta^{\nu \mu}_{\lambda_1 \lambda_2} }{\sqrt{5!}{\sqrt{2!}}} & 0
\end{array} \right)
\end{align*}
\item Type IIb decomposition with SL$(2)$-indices $k,l,m,... = 1,2$ and target-space indices $\underline{\kappa},\underline{\lambda},\underline{\mu},...=1,...,d-1$:
\begin{align*}
X^M &= \left(x^{\underline{\mu}},\tilde{x}^m_{\underline{\mu}},\frac{\tilde{x}_{{\underline{\mu}}_1 {\underline{\mu}}_2 {\underline{\mu}}_3}}{\sqrt{3!}},\frac{\tilde{x}^m_{{\underline{\mu}}_1 {\underline{\mu}}_2 {\underline{\mu}}_3 {\underline{\mu}}_4 {\underline{\mu}}_5}}{\sqrt{5!}} \right) \quad \in \mathcal{R}_1 \\
\mathcal{Q}^\mathcal{M} &= \left( Q_m, \frac{Q^{\underline{\mu}_1 {\underline{\mu}}_2}}{\sqrt{2!}} , \frac{Q^{{\underline{\mu}}_1 {\underline{\mu}}_2 {\underline{\mu}}_3 {\underline{\mu}}_4}_m}{\sqrt{4!}} , \frac{Q^{{\underline{\mu}}_1 {\underline{\mu}}_2 {\underline{\mu}}_3 {\underline{\mu}}_4 {\underline{\mu}}_5,{\underline{\mu}}}}{\sqrt{5!}} \right) \quad \in \mathcal{R}_2
\end{align*}

\begin{align*}
\eta_{\mathcal{M},KL}: \quad \left( \eta^m \right)_{KL} &= \left( \begin{array}{cccc} 0 & \delta^m_l \delta^{\underline{\lambda}}_{\underline{\kappa}} & 0 & 0 \\ \delta^m_k \delta^{\underline{\kappa}}_{\underline{\lambda}} & 0 & 0 & 0 \\ 0 & 0 & 0 & 0 \\ 0 & 0 & 0 & 0 \end{array} \right) \\ 
\left( \eta_{\underline{\mu}_1 \underline{\mu}_2} \right)_{KL} &= \left( \begin{array}{cccc} 0 & 0 & \frac{\delta^{\underline{\lambda}_1 \underline{\lambda}_2 \underline{\lambda}_3}_{ \underline{\mu}_1 \underline{\mu}_2 \underline{\kappa}}}{\sqrt{3!}} & 0 \\   0 & \epsilon_{kl} \delta^{\underline{\kappa}\underline{\lambda}}_{\underline{\mu}_1 \underline{\mu_2}} & 0 & 0 \\ \frac{\delta^{\underline{\lambda}_1 \underline{\lambda}_2 \underline{\lambda}_3}_{ \underline{\mu}_1 \underline{\mu}_2 \underline{\lambda}}}{\sqrt{3!}} & 0 & 0 & 0 \\ 0 & 0 & 0 & 0 \end{array} \right) \\
\left( \eta^m_{\underline{\mu}_1 \underline{\mu}_2 \underline{\mu}_3 \underline{\mu}_4} \right)_{KL} &= \left( \begin{array}{cccc} 0 & 0 & 0 & \frac{\delta^m_l \delta^{\underline{\lambda}_1 \underline{\lambda}_2 \underline{\lambda}_3 \underline{\lambda}_4 \underline{\lambda}_5}_{ \underline{\mu}_1 \underline{\mu}_2 \underline{\mu}_3 \underline{\mu}_4 \underline{\kappa}}}{\sqrt{5!}}  \\ 0 & 0 & \frac{\delta^m_k \delta^{\underline{\kappa} \underline{\lambda}_1 \underline{\lambda}_2 \underline{\lambda}_3}_{ \underline{\mu}_1 \underline{\mu}_2 \underline{\mu}_3 \underline{\mu}_4}}{\sqrt{3!}} & 0 \\ 0 & \frac{\delta^m_l \delta^{\underline{\lambda} \underline{\kappa}_1 \underline{\kappa}_2 \underline{\kappa}_3}_{ \underline{\mu}_1 \underline{\mu}_2 \underline{\mu}_3 \underline{\mu}_4}}{\sqrt{3!}} & 0 & 0 \\ \frac{\delta^m_k \delta^{\underline{\kappa}_1 \underline{\kappa}_2 \underline{\kappa}_3 \underline{\kappa}_4 \underline{\kappa}_5}_{ \underline{\mu}_1 \underline{\mu}_2 \underline{\mu}_3 \underline{\mu}_4 \underline{\lambda}}}{\sqrt{5!}} & 0 & 0 & 0 \end{array} \right) \\
\left( \eta^m_{{\underline{\mu}}_1 {\underline{\mu}}_2 {\underline{\mu}}_3 {\underline{\mu}}_4 {\underline{\mu}}_5,{\underline{\mu}}} \right)_{KL} &= \left( \begin{array}{cccc} 0 & 0 & 0 & 0  \\ 0 & 0 & 0 & \epsilon^{lk} \frac{\delta_{\underline{\mu}}^{\underline{\kappa}} \delta^{\underline{\lambda}_1 \underline{\lambda}_2 \underline{\lambda}_3 \underline{\lambda}_4 \underline{\lambda}_5}_{ \underline{\mu}_1 \underline{\mu}_2 \underline{\mu}_3 \underline{\mu}_4 \underline{\mu}_5}}{\sqrt{5!}} \\ 0 & 0 & \frac{1}{2} \left( \frac{\delta^{ \underline{\kappa}_1 \underline{\kappa}_2 \underline{\kappa}_3 \underline{\nu}_1 \underline{\nu}_2} _{\underline{\mu}_1 \underline{\mu}_2 \underline{\mu}_3 \underline{\mu}_4 \underline{\mu_5}} \delta_{\underline{\nu}_1 \underline{\nu}_2 \underline{\mu}}^{\underline{\lambda}_1 \underline{\lambda}_2 \underline{\lambda_3}}}{2!\sqrt{3!3!}} + (\underline{\kappa} \leftrightarrow \underline{\lambda})\right)  & 0 \\ 0 & \epsilon^{kl} \frac{\delta_{\underline{\mu}}^{\underline{\lambda}} \delta^{\underline{\kappa}_1 \underline{\kappa}_2 \underline{\kappa}_3 \underline{\kappa}_4 \underline{\kappa}_5}_{ \underline{\mu}_1 \underline{\mu}_2 \underline{\mu}_3 \underline{\mu}_4 \underline{\mu}_5}}{\sqrt{5!}} & 0 & 0 \end{array} \right)
\end{align*}
The $\omega$-symbols are given by the same terms as the $\eta$-symbols, but with different signs similar to the M-theory section:
\begin{align*}
\omega_{\bullet,MN} = \left( \begin{array}{cccc}  0 & - & - & - \\
+ & - & - & - \\ + & - & - & 0 \\ + & - & 0 & 0 \end{array} \right).
\end{align*}

\end{itemize}
\pagebreak

\bibliographystyle{jhep}
\bibliography{References}
\end{document}